\documentclass[acmlarge]{acmart}

\AtBeginDocument{%
  \providecommand\BibTeX{{%
    \normalfont B\kern-0.5em{\scshape i\kern-0.25em b}\kern-0.8em\TeX}}}

\setcopyright{acmcopyright}
\copyrightyear{2018}
\acmYear{2018}
\acmDOI{10.1145/3548455}

\usepackage{graphicx}
\usepackage{amsmath}
\usepackage{amsthm}
\usepackage{booktabs}
\usepackage{algorithm}
\usepackage{algorithmic}
\usepackage{subfigure}
\usepackage{multirow}
\usepackage{array}
\usepackage{makecell}
\usepackage{enumerate}
\usepackage{color}
\usepackage[justification=centering]{caption}
\usepackage{diagbox}

\acmJournal{TOIS}
\acmVolume{37}
\acmNumber{4}
\acmArticle{111}
\acmMonth{8}

\begin{document}

\title{A Survey on Cross-domain Recommendation: Taxonomies, Methods, and Future Directions}

\author{Tianzi Zang}
\affiliation{%
  \institution{Shanghai Jiao Tong University}
  \streetaddress{No. 800 Dongchuan Road, Minhang District, Shanghai}
  \city{Shanghai}
  \country{China}
  \postcode{200240}
  }
\email{zangtianzi@sjtu.edu.cn}

\author{Yanmin Zhu}
\affiliation{%
  \institution{Shanghai Jiao Tong University}
  \streetaddress{No. 800 Dongchuan Road, Minhang District, Shanghai}
  \city{Shanghai}
  \country{China}
  \postcode{200240}
  }
\email{yzhu@sjtu.edu.cn}

\author{Haobing Liu}
\affiliation{%
  \institution{Shanghai Jiao Tong University}
  \streetaddress{No. 800 Dongchuan Road, Minhang District, Shanghai}
  \city{Shanghai}
  \country{China}
  \postcode{200240}
  }
\email{liuhaobing@sjtu.edu.cn}

\author{Ruohan Zhang}
\affiliation{%
  \institution{Shanghai Jiao Tong University}
  \streetaddress{No. 800 Dongchuan Road, Minhang District, Shanghai}
  \city{Shanghai}
  \country{China}
  \postcode{200240}
  }
\email{Zhangruohan@sjtu.edu.cn}


\author{Jiadi Yu}
\affiliation{%
  \institution{Shanghai Jiao Tong University}
  \streetaddress{No. 800 Dongchuan Road, Minhang District, Shanghai}
  \city{Shanghai}
  \country{China}
  \postcode{200240}
  }
\email{jiadiyu@sjtu.edu.cn}


\renewcommand{\shortauthors}{Zang et al.}

\begin{abstract}
Traditional recommendation systems are faced with two long-standing obstacles, namely, data sparsity and cold-start problems, which promote the emergence and development of Cross-Domain Recommendation (CDR). The core idea of CDR is to leverage information collected from other domains to alleviate the two problems in one domain. Over the last decade, many efforts have been engaged for cross-domain recommendation. Recently, with the development of deep learning and neural networks, a large number of methods have emerged. However, there is a limited number of systematic surveys on CDR, especially regarding the latest proposed methods as well as the recommendation scenarios and recommendation tasks they address. In this survey paper, we first proposed a two-level taxonomy of cross-domain recommendation which classifies different recommendation scenarios and recommendation tasks. We then introduce and summarize existing cross-domain recommendation approaches under different recommendation scenarios in a structured manner. We also organize datasets commonly used. We conclude this survey by providing several potential research directions about this field.
\end{abstract}

\ccsdesc[500]{Information systems~Recommender systems}

\keywords{cross-domain recommendation, survey, datasets, machine learning, deep learning}
\maketitle

\section{Introduction}
In the era of information explosion, people are easy to be overwhelmed by a large amount of information. Millions of new products, text, and videos are being released all the time, which makes it difficult for people to find their interested items. Users' historical interactions are believed to contain rich information about user interests that can be used to predict their future interests. This promotes the emergence of recommendation systems (RS). The basic idea of RS is to analyze and estimate users' interests, and then select items that users may be interested in from a large number of candidate items and recommend them to the users. 

Although recommendation systems have been proved to play a significant role in a variety of applications, there are two long-standing obstacles that greatly limit the performance of recommendation systems. On the one hand, the number of user-item interaction records often tends to be small and is insufficient to mine user interests well, which is called the data sparsity problem. On the other hand, for any service, there are constantly new users joining, for whom there are no historical interaction records. Traditional recommendation systems cannot make recommendations to these users, which is called the cold-start problem. As more and more users begin to interact with more than one domains (e.g., music and book), it increases opportunities of leveraging information collected from other domains to alleviate the two problems (i.e., data sparsity and cold-start problems) in one domain. This idea leads to Cross-Domain Recommendation (CDR) which has attracted increasing attention in recent years.

Compared with traditional recommendation systems, cross-domain recommendation is more complicated. First, considering the relations between user sets and item sets of two domains, there are different recommendation scenarios of cross-domain recommendation such as user overlap or non-overlap~\cite{DBLP:conf/icdm/CremonesiTT11,DBLP:journals/csur/KhanIG17}. Second, the recommendation tasks of cross-domain recommendation are various. For example. the recommended items and the user may be in the same domain or different domains. The goal of recommendation may be improving performance in one specific domain or multiple domains. Third, traditional recommendation systems only need to focus on how to model user interests from historical interaction records. For cross-domain recommendation, besides modeling user interests within a domain, it also needs to consider how to transfer the knowledge (i.e., user interests) among domains. This leads to two core issues for cross-domain recommendation, namely, \emph{what to transfer} and \emph{how to transfer}. What to transfer is how to mine useful knowledge in each domain, and how to transfer focuses on how to establish linkages between domains and realize the transfer of knowledge.

Over the last decade or so, many efforts have been engaged for cross-domain recommendation. To answer the question of what to transfer, existing studies are dedicated to applying different methods to extract useful knowledge in each domain. Traditional machine learning methods, such as matrix factorization~\cite{DBLP:conf/kdd/SinghG08,DBLP:conf/ijcai/XinLLHWG15}, factorization machines~\cite{DBLP:conf/aaai/LiDL19,DBLP:conf/ecir/LoniSLH14}, co-clustering~\cite{DBLP:conf/cikm/MorenoSRS12,DBLP:conf/ijcai/LiYX09,DBLP:conf/wsdm/WangFGCH19}, and latent semantic analysis~\cite{DBLP:conf/sdm/LuPXYZZ13,DBLP:journals/ijon/TanBQCC14} have been widely applied.
In recent years, with the emergence and development of deep learning technologies, many approaches based on deep learning have been proposed~\cite{DBLP:conf/www/ElkahkySH15,DBLP:conf/ijcai/YanCGLJ19,DBLP:conf/cikm/ZhaoLF19,DBLP:conf/cikm/HuZY18,DBLP:conf/sigir/MaRLCMR19,hu2018mtnet,DBLP:conf/www/GaoCFZ00J19}, which greatly improves the accuracy and performance of cross-domain recommendation.
To answer the question of how to transfer, a straightforward idea is to utilize overlapping entities, either users or items, to directly establish relationships between domains~\cite{DBLP:conf/cikm/ZhuC0LZ19,DBLP:conf/ijcai/ZhuWCLZ20,DBLP:conf/aaai/PereraZ20,DBLP:conf/kdd/SinghG08,DBLP:conf/www/HuCXCGZ13,DBLP:conf/ijcai/YanCGLJ19,DBLP:conf/sigir/ChenZWMLLM19}. 
When there are no overlapping entities, some efforts are also made to establish linkages by extracting cluster-level patterns~\cite{DBLP:conf/ijcai/LiYX09,DBLP:conf/cikm/MorenoSRS12,DBLP:conf/wsdm/ShuWTWL18,DBLP:conf/wsdm/WangFGCH19} or resorting to other auxiliary information (e.g., users' generated tags, reviews, user profiles, and item content)~\cite{
DBLP:conf/recsys/Fernandez-TobiasC14,DBLP:conf/um/ShiLH11,DBLP:conf/cikm/YangHQXW15,DBLP:conf/ijcnn/ZhangH0019}.

In the literature, there are several surveys on cross-domain recommendation. 
Li et al.~\cite{DBLP:conf/ictai/Li11a} first gave a brief survey in which they proposed that there were three different types of domains, that is, system domains, data domains, and temporal domains. To be more specific, for different system domains, there are different types of items (e.g., books and movies) or items of different genres (e.g., comedy movies and fiction movies). For data domains, it refers to the fact that users' preferences towards items can be stored in multiple data types (e.g., explicit numeric rating data and implicit binary feedback data) and each type of data is treated as a domain. For the temporal domain, the interaction records are divided into several temporal slices according to timestamps and each time slice constitutes a domain. This classification of domain types was found to be widely cited by later researchers. Cremonesi et al.~\cite{DBLP:conf/icdm/CremonesiTT11} identified four different cross-domain scenarios based on the relations between user sets and items sets of two domains, that is, no overlap, user overlap, item overlap, and full overlap, which have been recognized by the following studies~\cite{fernandez2012cross,DBLP:journals/csur/KhanIG17}.
Ricci et al. wrote the ``Recommender System Handbook"~\cite{DBLP:reference/sp/2015rsh} and chapter 27 carefully introduced cross-domain recommendation research by 2015.
Khan et al.~\cite{DBLP:journals/csur/KhanIG17} then wrote another survey in which they made a detailed comparison and discussion of previous surveys and identified domain type, user-item overlap scenario, and recommendation tasks as three building blocks of cross-domain recommender systems. They also made a detailed summary and analysis of enabling algorithms, identified problems, and future directions. 
The most recent survey was written by Zhu et al.~\cite{DBLP:conf/ijcai/ZhuW00L021} in which they gave definitions to four different cross-domain recommendation tasks in detail, that is, single-target CDR, single-target MDR (multi-domain recommendation), dual-target CDR, and multi-target CDR. They present challenges for different tasks and categorize existing approaches for single-target CDR.

Since most of existing surveys on cross-domain recommendation have been published for several years, we found that they can not meet the current demand for research in this area. We write this survey mainly for the following three reasons. 
First, the classification of recommendation scenarios and recommendation tasks in existing surveys are coarse-grained. Considering the actual situation of cross-domain recommendation research, the classification can be further refined (see section~\ref{recommendation_scenario} and section~\ref{recommendation_task} for details). For example, existing surveys just classified the relations between user sets and item sets into two cases: overlap and non-overlap. We propose that the overlap relation can be further divided into partial overlap and full overlap. 
Second, many significant advances in cross-domain recommendation have happened after these surveys were published. In particular, the rising popularity and rapid development of deep learning-based technologies in recent years have affected the field of cross-domain recommendation to a large extend. Although the recent survey paper~\cite{DBLP:conf/ijcai/ZhuW00L021} mentioned some new approaches, it only classified approaches that are for single-target CDR.
Therefore, there is still a need for a survey that summarizes the most recent approaches in cross-domain recommendation in detail. 
Third, since there exist different recommendation scenarios and recommendation tasks in cross-domain recommendation, a method proposed in one scenario is often not applicable to another scenario. 
Therefore, when discussing existing studies, it is essential to categorize them according to the classification of recommendation scenarios and tasks, which has not been well addressed by existing surveys.

For a literature survey, it is essential for finding every relevant piece of work. We adopted a hybrid approach for searching the relevant literature. We first used Google Scholar as the main search engine to discover related papers. We then screened most of the related high-profile conferences such as NIPS, ICML, ICLR, SIGKDD, WWW, AAAI, SIGIR, IJCAI, WSDM, and RecSys, to find out the recent works. The major keywords we use include ``cross-domain rec'', ``cross-system rec'', ``cross-network rec'', and ``cross-platform rec''. We also pay attention to the references mentioned in the relevant work section of each paper to prevent omissions of relevant literature.


The contributions of this paper are summarized as follows:

\begin{itemize}
\item We propose an original two-level taxonomy of cross-domain recommendation which identifies $9$ different recommendation scenarios based on the overlap of user/item sets and $4$ different recommendation tasks.
\item We summarize existing research works and make a method-based categorization of them concerning the recommendation scenarios and recommendation tasks they address.
\item We introduce frequently used datasets in cross-domain recommendation, group them into different categories and explain how they can be used in researches.
\item We outline further potential research directions of cross-domain recommendation.
\end{itemize}

The remainder of this paper is organized as follows. In section~\ref{taxonomy}, we introduce the notations adopted in this paper and our proposed two-level taxonomy about cross-domain recommendation scenarios and recommendation tasks. In section~\ref{scenario_1}$-$\ref{scenario_3}, we, respectively, summarize and make method-based categorization of existing cross-domain works under three widely studied recommendation scenarios in a structured manner. Section~\ref{dataset} will introduce datasets frequently used in cross-domain recommendation. In section~\ref{challenge_future_direction}, we present potential future search directions. Finally, we conclude this paper in Section~\ref{conclusion}.

\section{taxonomy}
\label{taxonomy}
In this section, we first introduce the related notations adopted in this paper. Then we introduce our proposed two-level taxonomy of recommendation scenarios and recommendation tasks, followed by the method-based categorization of existing researches under different recommendation scenarios. 

\subsection{Notations}
Without loss of generality, we consider the cross-domain recommendation when only two domains $D^{A}$ and $D^{B}$ are involved. The notations introduced here can be easily extended to situations with multiple domains. $\mathcal{U}^A$ and $\mathcal{I}^A$, respectively, denote the user set and item set in $D^{A}$ while $\mathcal{U}^B$ and $\mathcal{I}^B$ are the user set and item set in $D^{B}$. Two matrices $\textbf{R}^{A}\in \mathbb{R}^{m^A\times n^A}$ and $\textbf{R}^{B}\in \mathbb{R}^{m^B\times n^B}$ represent interactions between users and items in each domain, where $m^A=|\mathcal{U}^A|, n^A=|\mathcal{I}^A|, m^B=|\mathcal{U}^B|$ and $n^B=|\mathcal{I}^B|$ denote the number of users and items in $D^{A}$ and $D^{B}$. Each element $r_{mn}\in \textbf{R}^A$ (or $\textbf{R}^B$) can be numeric denoting a user's explicit rating to an item or binary number representing an implicit interaction (e.g., click, purchase, add to cart) between a user-item pair. In addition, in each domain, there may be another two matrices $\textbf{X}^A\in\mathbb{R}^{m^A\times p^A}, \textbf{Y}^A\in\mathbb{R}^{n^A\times q^A}$ ($\textbf{X}^B\in\mathbb{R}^{m^B\times p^B}, \textbf{Y}^B\in\mathbb{R}^{n^B\times q^B}$ in $D^{B}$) representing user-profiles and items attributes, respectively.



\begin{figure*}[t]
    \centering
    \includegraphics[width=\textwidth]{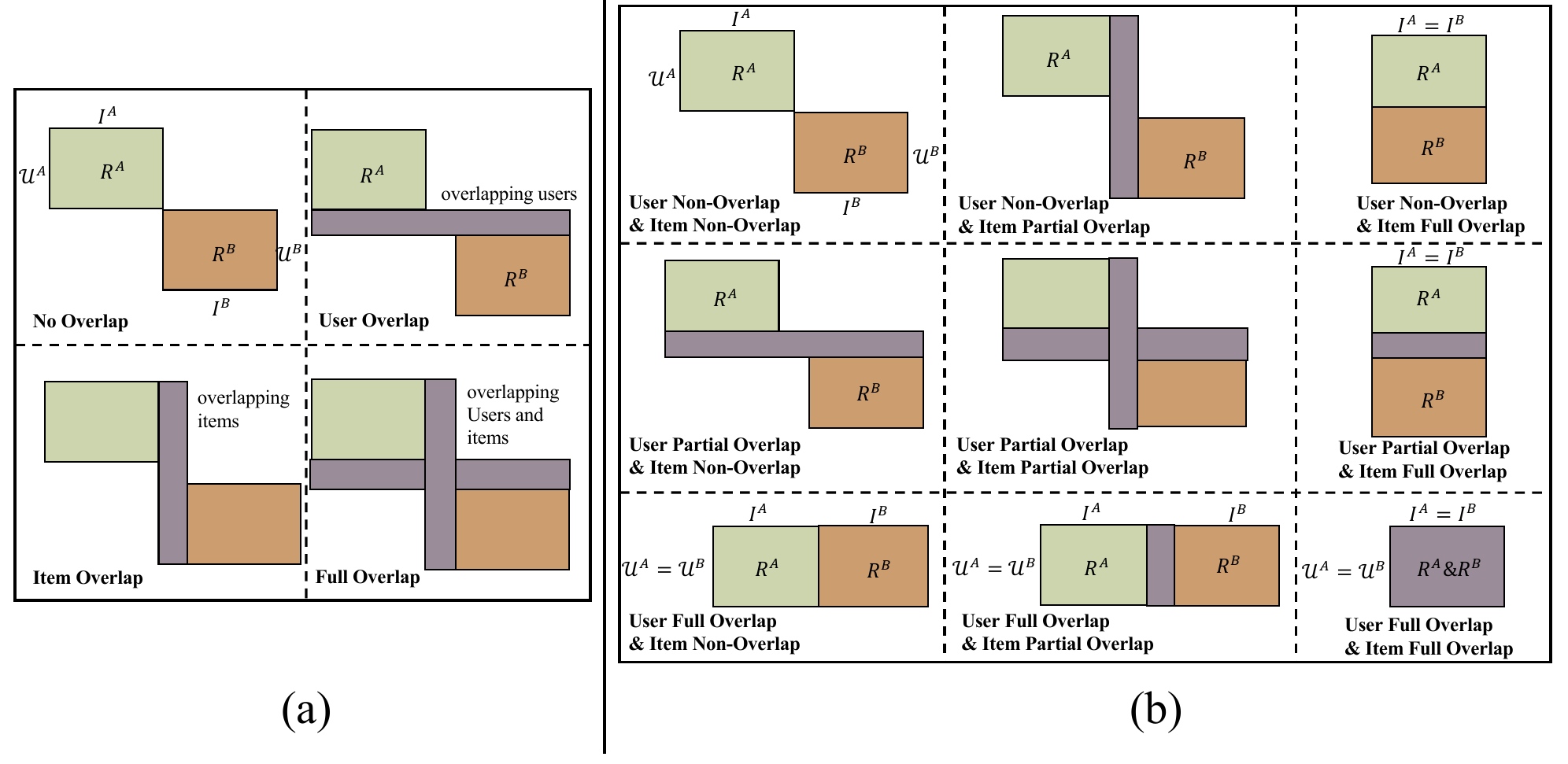}
    \caption{Comparison between previous classification of recommendation scenarios and ours. (a) illustration of scenarios identified by previous surveys. (b) illustration of scenarios identified in this paper.}
    \label{illustration_of_recommendation_scenarios}
\end{figure*}

\subsection{The Proposed Two-level Taxonomy}
\label{two_level_taxonomy}
Different from traditional single-domain recommendation systems, cross-domain recommendation is more complicated as there exist different recommendation scenarios and recommendation tasks. Therefore, we first proposed a two-level taxonomy, each level of which consists of two dimensions to, respectively, classify cross-domain recommendation scenarios and recommendation tasks. With our proposed taxonomy, all of the current studies can be classified and positioned, which can help researchers quickly understand the characteristics of each study. In the following, we introduce our proposed two-level taxonomy in detail.

\subsubsection{The First Level: Classification of recommendation scenarios}
\label{recommendation_scenario}
The first level of the proposed taxonomy is about recommendation scenarios and the classification criteria is the relations between user sets and item sets of two domains. To be more specific, there are two dimensions, that is, the overlap of user sets and the overlap of item sets.  

The classification of recommendation scenarios was first proposed by Cremonesi et al.~\cite{DBLP:conf/icdm/CremonesiTT11} in which they identified four different recommendation scenarios (shown in Fig.\ref{illustration_of_recommendation_scenarios} (a)), that is, no overlap, user overlap, item overlap, and full overlap. This classification was widely accepted by subsequent researchers. However, through analysis and summary of recent works, we propose that this classification is coarse-trained and can be further refined. Still considering the relations between user sets and item sets of two domains, we describe these two dimensions in detail as follows.

\noindent\textbf{Dimension 1: the overlap of user sets.}
This dimension refers to whether there are common users between the user sets of two domains. Previous studies only divided user sets into overlap and non-overlap, but we propose that overlap can be further subdivided into partial overlap and full overlap. Therefore, under this dimension, cross-domain recommendation can be divided into three sub-categories. The first category is \emph{user non-overlap}. Following the notations defined above, it can be denoted as $\mathcal{U}^{A} \cap \mathcal{U}^{B} = \varnothing$. The second category is \emph{user partial overlap} which means $\mathcal{U}^{A}\neq\mathcal{U}^{B}$ and $\mathcal{U}^{A}\cap\mathcal{U}^{B}\neq\varnothing$. The third category is \emph{user full overlap} which means $\mathcal{U}^{A} = \mathcal{U}^{B}$.

\noindent\textbf{Dimension 2: the overlap of item sets.}
This dimension refers to whether there are common items between the item sets of two domains. Previous studies also only divided items into overlap and non-overlap categories. Similar to the first dimension, considering the overlap of item sets, cross-domain recommendation can also be divided into three sub-categories, that is, \emph{item non-overlap}, \emph{item partial overlap} and \emph{item full overlap}. Similarly, they can be denoted, respectively, as $\mathcal{I}^{A} \cap \mathcal{I}^{B} = \varnothing$, $\mathcal{I}^{A}\neq\mathcal{I}^{B}$ and $\mathcal{I}^{A}\cap\mathcal{I}^{B}\neq\varnothing$, and $\mathcal{I}^{A} = \mathcal{I}^{B}$.

\begin{table}[t]
	\caption{Categorization of existing works according to recommendation scenarios.}
	\label{problem_scenario}
	\centering
	\setlength{\tabcolsep}{2.2 mm}
	\begin{tabular}{c|c|c}
	\toprule
	\textbf{Overlap of user sets}& \textbf{Overlap of item sets} & \textbf{Existing works}\\
	\cline{1-3}
	\multirow{4}{*}{User non-overlap} & \multirow{2}{*}{Item non-overlap} &~\cite{DBLP:conf/ijcai/LiYX09,DBLP:conf/cikm/MorenoSRS12,DBLP:conf/pkdd/GaoLCLGG13,DBLP:conf/aaai/RenGLG15,DBLP:conf/wsdm/HeZYY18,DBLP:journals/dss/ZhangWLLZ17,DBLP:conf/kdd/ChenHL13,DBLP:conf/iconip/ZhangWLZ18,DBLP:conf/wsdm/ShuWTWL18,DBLP:conf/icml/LiYX09,DBLP:conf/aistats/IwataK15,DBLP:conf/wsdm/WangFGCH19}\\
	& & ~\cite{DBLP:conf/ecweb/EnrichBR13,DBLP:conf/recsys/Fernandez-TobiasC14,DBLP:conf/um/ShiLH11,DBLP:conf/fuzzIEEE/HaoZL16,DBLP:conf/ijcnn/ZhangH0019,DBLP:conf/icdm/FangGLLL15,DBLP:conf/cidm/KumarKHCA14,DBLP:conf/cikm/YangHQXW15,DBLP:conf/aaai/ZhaoPXZLY13,DBLP:journals/ai/ZhaoPY17,DBLP:conf/aaai/ZhangJLD016}\\
	\cline{2-3}
	 & Item partial overlap & symmetric to user partial overlap $\&$ item non-overlap\\
	 \cline{2-3}
	 & Item full overlap & symmetric to user full overlap $\&$ item non-overlap\\
	 \cline{1-3}
	\multirow{4}{*}{User partial overlap} & \multirow{2}{*}{Item non-overlap} & ~\cite{DBLP:conf/pkdd/RafailidisC16,DBLP:conf/cikm/RafailidisC17,DBLP:journals/tnn/ZhangLWZ19,DBLP:conf/sigir/YangYYLC17,DBLP:conf/cikm/ZhuC0LZ19,DBLP:conf/ijcai/ZhuWCLZ20,DBLP:conf/mm/PereraZ17,DBLP:conf/aaai/PereraZ20,DBLP:conf/recsys/CuiWZZ20,DBLP:conf/dasfaa/LiPWXYH20,DBLP:conf/aaai/JiangCYXY16,DBLP:journals/corr/abs-2104-08490,DBLP:conf/cikm/XuXCZ21}\\
	& & ~\cite{DBLP:conf/ijcai/ManSJC17,DBLP:conf/dasfaa/WangPWYFH18,DBLP:conf/aaai/FuPWXL19,DBLP:conf/sigir/BiSYWWX20a,DBLP:conf/sigir/BiSYWWX20,DBLP:conf/ijcai/ZhuWCLOW18,DBLP:conf/cikm/KangHLY19,DBLP:conf/ijcai/ZhangLHMCLT20,DBLP:conf/sigir/ZhaoLXDS20,DBLP:conf/sigir/Wang0NC17,DBLP:journals/corr/abs-2108-07976,DBLP:conf/sigir/ZhuGZXXZL021,DBLP:conf/wsdm/ZhuTLZXZLH22,DBLP:conf/cikm/WangZZWZH21}\\
	\cline{2-3}
	 & Item partial overlap& -\\
	 \cline{2-3}
	 & Item full overlap& -\\
	 \cline{1-3}
	\multirow{4}{*}{User full overlap} & \multirow{2}{*}{Item non-overlap} & \cite{DBLP:conf/sdm/LuPXYZZ13,DBLP:journals/ijon/TanBQCC14,DBLP:conf/recsys/SahebiBB17,DBLP:conf/kdd/SinghG08,DBLP:conf/cikm/MaYLK08a,DBLP:conf/ijcai/XinLLHWG15,DBLP:conf/dasfaa/ZhaoHWH18,DBLP:journals/ijon/HuangZWHC19,DBLP:conf/www/HuCXCGZ13,DBLP:conf/icml/LiuHW15,DBLP:conf/dasfaa/SongPWFHY17,DBLP:conf/ecir/LoniSLH14,DBLP:conf/aaai/LiDL19,DBLP:conf/ijcai/MaZWLLM18,DBLP:conf/ijcai/0008T0ZNY21,DBLP:conf/wsdm/ZhaoYY22}\\
	&&\cite{DBLP:conf/www/ElkahkySH15,DBLP:conf/www/LianZXS17,DBLP:conf/icdm/HeLZLNH18,DBLP:conf/cikm/ZhaoLF19,DBLP:conf/ijcai/YanCGLJ19,DBLP:conf/sigir/ChenZWMLLM19,DBLP:conf/cikm/HuZY18,DBLP:conf/www/LiuZZLSX0X20,DBLP:conf/wsdm/0008T20,DBLP:conf/sigir/MaRLCMR19,DBLP:conf/cikm/LiuLLP20,hu2018mtnet,DBLP:conf/www/GaoCFZ00J19,DBLP:conf/ijcai/PereraZ18,DBLP:journals/corr/abs-2202-04893}\\
	\cline{2-3}
	 & Item partial overlap & -\\
	 \cline{2-3}
	 & Item full overlap& equivalent to single-domain recommendation\\
	\bottomrule
	\end{tabular}
\end{table}

With these two dimensions, we extend the classification of recommendation scenarios from previous $4$ categories to $9$ categories and show the comparison in Fig.~\ref{illustration_of_recommendation_scenarios}. We also present the categorization of existing works according to recommendation scenarios which is shown in Table~\ref{problem_scenario}.
It is worth noting that there are currently no studies on three scenarios (i.e., user partial overlap $\&$ item partial overlap, user partial overlap $\&$ item full overlap, and user full overlap $\&$ item partial overlap). To ensure the integrity of the tree-structured categorization, we retain these nodes. It should also be pointed out that the recommendation scenario with user non-overlap $\&$ item partial overlap is symmetric to the case with user partial overlap $\&$ item non-overlap. Similarly, the scenario with user non-overlap $\&$ item full overlap is symmetric to the case with user full overlap $\&$ item non-overlap.
Most of the approaches proposed for one scenario apply to the other scenario by exchanging users and items. Therefore, we will not discuss in detail the scenarios with user non-overlap $\&$ item partial overlap and user non-overlap $\&$ item full overlap.
Moreover, the recommendation scenario with user full overlap $\&$ item full overlap is equivalent to single-domain recommendation and is therefore outside the scope of this paper. 

\subsubsection{The Second Level: Classification of Recommendation Tasks}
\label{recommendation_task}
The second level of our proposed taxonomy is about recommendation tasks which also consists of two dimensions. Based on whether the recommended items are in the same domain as the user, cross-domain recommendation can be divided into two categories, that is, \emph{intra-domain recommendation} and \emph{inter-domain recommendation}. Considering whether the number of target domains is single or multiple, cross-domain recommendation can be divided into other two categories, i.e., \emph{single-target cross-domain recommendation} and \emph{multi-target cross-domain recommendation}. For the convenience of understanding, we elaborate on these two dimensions.

\noindent\textbf{Dimension 1: whether recommended items are in the same domain as the user.} Considering this dimension, there are two different cross-domain recommendation tasks, that is, intra-domain recommendation and inter-domain recommendation. For \emph{intra-domain recommendation}, the recommended items are in the same domain as the user which means we recommend a subset of items $\mathcal{I}_u\subseteq\mathcal{I}^A$ to a user $u\in\mathcal{U}^A$ or we recommend $\mathcal{I}_u\subseteq\mathcal{I}^B$ to a user $u\in\mathcal{U}^B$. For \emph{inter-domain recommendation}, the recommended items are from a different domain with the user, that is, we recommend $\mathcal{I}_u\subseteq\mathcal{I}^A$ to a user $u\in\mathcal{U}^B$ or $\mathcal{I}_u\subseteq\mathcal{I}^B$ to a user $u\in\mathcal{U}^A$. Specifically, inter-domain recommendation is often referred to as cold-start user recommendation which is because that $u\in \mathcal{U}^A$ (or $\mathcal{U}^B$) have no interaction with items in $\mathcal{I}^B$ (or $\mathcal{I}^A$) and, therefore, can be seen as a cold-start user in domain $A$ (or domain $B$).

\noindent\textbf{Dimension 2: whether the number of target domains is single or multiple.} Based on this dimension, there are another two cross-domain recommendation tasks, that is, single-target recommendation and multi-target recommendation. For \emph{single-target recommendation}, the domain with a denser interaction matrix is generally treated as the auxiliary/source domain and the other domain with a sparser interaction matrix is referred to as the target domain. The goal of the recommendation is to improve the recommendation performance in the target domain. For \emph{multi-target recommendation}, it is based on the assumption that the interaction matrices of both domains are sparse or both domains are sparse in some kinds of information, thus, the performance of both domains can be improved by utilizing knowledge from the other domain. There is no source-target distinction and the aim is to simultaneously enrich both domains and improve the performance of both domains.

\begin{table*}[t]
	\caption{Categorization of existing works according to recommendation tasks.}
	\label{recommendation_tasks}
	\begin{tabular}{|c|c|c|}
	\hline
	\diagbox{Dimension 2}{Dimension 1} & Intra-domain recommendation 	& Inter-domain recommendation\\
	\hline
	\multirow{2}{*}{\shortstack{Single-target\\recommendation}} & \cite{DBLP:conf/ijcai/LiYX09,DBLP:conf/cikm/MorenoSRS12,DBLP:conf/wsdm/HeZYY18,DBLP:conf/kdd/ChenHL13,DBLP:conf/iconip/ZhangWLZ18,DBLP:conf/um/ShiLH11,DBLP:conf/wsdm/ShuWTWL18,DBLP:conf/mm/PereraZ17,DBLP:conf/aaai/PereraZ20,DBLP:conf/ecweb/EnrichBR13} & \cite{DBLP:conf/cidm/KumarKHCA14,DBLP:conf/cikm/YangHQXW15,DBLP:conf/sigir/Wang0NC17,DBLP:conf/ijcai/ManSJC17,DBLP:conf/dasfaa/WangPWYFH18,DBLP:conf/aaai/FuPWXL19,DBLP:conf/sigir/BiSYWWX20a,DBLP:conf/sigir/BiSYWWX20,DBLP:conf/ijcai/ZhuWCLOW18}\\
	& \cite{DBLP:conf/recsys/Fernandez-TobiasC14,DBLP:conf/aaai/ZhaoPXZLY13,DBLP:journals/ai/ZhaoPY17,DBLP:conf/ecir/LoniSLH14,DBLP:conf/aaai/LiDL19,hu2018mtnet,DBLP:conf/www/GaoCFZ00J19,DBLP:conf/ijcai/MaZWLLM18,DBLP:conf/ijcai/PereraZ18,DBLP:journals/corr/abs-2202-04893} &~\cite{DBLP:conf/sigir/ZhuGZXXZL021,DBLP:conf/wsdm/ZhuTLZXZLH22,DBLP:conf/cikm/KangHLY19,DBLP:conf/cikm/WangZZWZH21}\\
	\hline
	\multirow{4}{*}{\shortstack{Multi-target\\recommendation}} & \cite{DBLP:conf/pkdd/GaoLCLGG13,DBLP:conf/aaai/RenGLG15,DBLP:conf/icml/LiYX09,DBLP:conf/aistats/IwataK15,DBLP:conf/wsdm/WangFGCH19,
DBLP:conf/um/ShiLH11,DBLP:conf/icdm/FangGLLL15,DBLP:conf/fuzzIEEE/HaoZL16,DBLP:conf/ijcnn/ZhangH0019,DBLP:conf/aaai/ZhangJLD016,DBLP:conf/ijcai/0008T0ZNY21,DBLP:conf/wsdm/ZhaoYY22} &  \\
	&\cite{DBLP:conf/recsys/CuiWZZ20,DBLP:conf/ijcai/ZhangLHMCLT20,DBLP:conf/ijcai/ZhuWCLZ20,DBLP:conf/cikm/ZhuC0LZ19,DBLP:conf/aaai/JiangCYXY16,DBLP:conf/pkdd/RafailidisC16,DBLP:conf/cikm/RafailidisC17,DBLP:journals/tnn/ZhangLWZ19,DBLP:conf/sigir/YangYYLC17}&\cite{DBLP:conf/dasfaa/LiPWXYH20,DBLP:conf/sigir/ZhaoLXDS20}\\
	& \cite{DBLP:conf/kdd/SinghG08,DBLP:conf/cikm/MaYLK08a,DBLP:conf/ijcai/XinLLHWG15,DBLP:conf/dasfaa/ZhaoHWH18,DBLP:conf/www/HuCXCGZ13,DBLP:conf/icml/LiuHW15,DBLP:conf/dasfaa/SongPWFHY17,DBLP:conf/www/ElkahkySH15,DBLP:conf/www/LianZXS17,DBLP:conf/icdm/HeLZLNH18,DBLP:conf/cikm/XuXCZ21}& \\
	&\cite{DBLP:conf/cikm/ZhaoLF19,DBLP:conf/ijcai/YanCGLJ19,DBLP:conf/sigir/ChenZWMLLM19,DBLP:conf/cikm/HuZY18,DBLP:conf/dasfaa/WangPWYFH18,DBLP:conf/www/LiuZZLSX0X20,DBLP:conf/wsdm/0008T20,DBLP:conf/sigir/MaRLCMR19,DBLP:conf/cikm/LiuLLP20,DBLP:journals/corr/abs-2202-04920,DBLP:journals/corr/abs-2108-07976,DBLP:journals/corr/abs-2104-08490}&\\
	\hline
	\end{tabular}
\end{table*}

Table~\ref{recommendation_tasks} shows the categorization of representative cross-domain recommendation studies according to recommendation tasks. As we can see, the most widely studied task is the intra-domain multi-target recommendation followed by the intra-domain single-target recommendation. The inter-domain multi-task recommendation is the least studied, with only two corresponding papers.

\begin{figure*}[t]
    \centering
    \includegraphics[width=\textwidth]{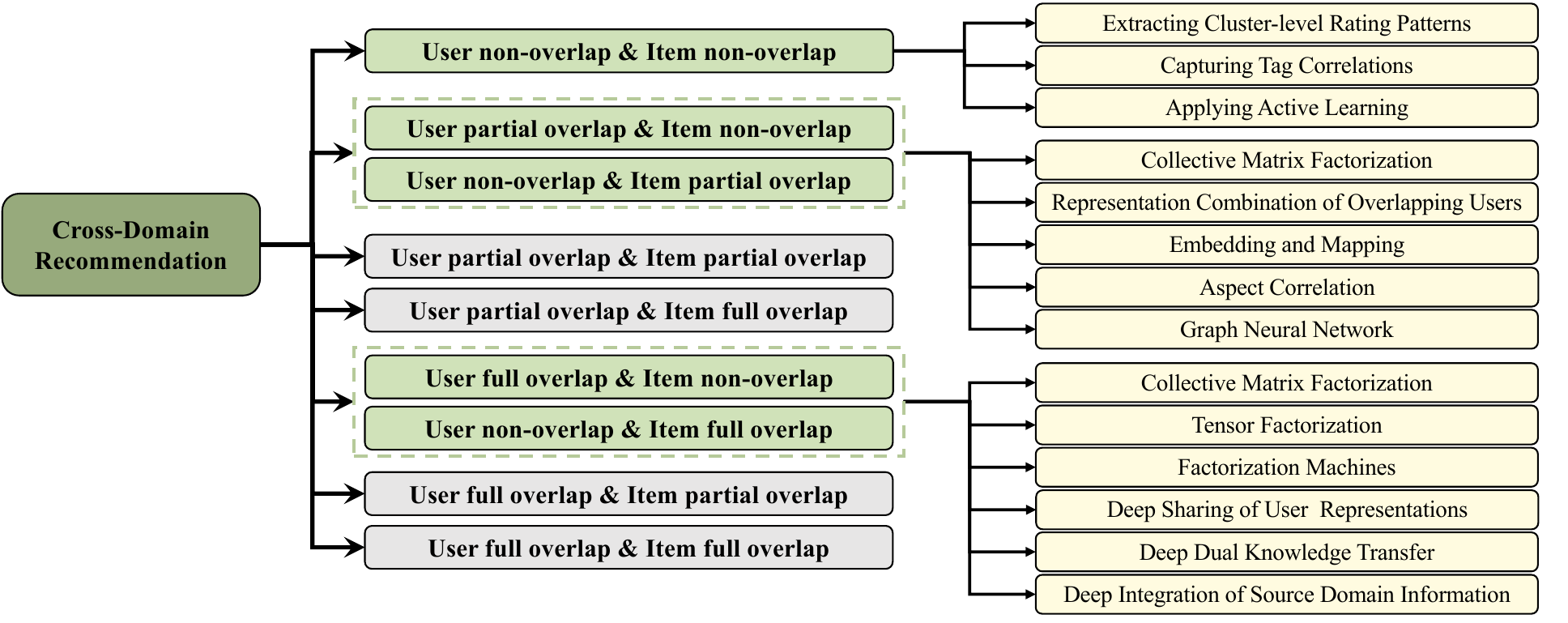}
    \caption{Method-based categorization under different recommendation scenarios.}
    \label{overview_of_methods}
\end{figure*}

\subsection{Method-based Categorization under Different Recommendation Scenarios}

Based on our proposed two-level taxonomy, we have identified $9$ different recommendation scenarios and $4$ different recommendation tasks. In the following, we will systematically review and analyze the approaches of existing works which is the core of this paper.
However, approaches proposed under a scenario may not apply to another scenario. It is meaningless to mix approaches for introduction regardless of which scenario they are aimed at. Therefore, before introducing existing studies, we first make a method-based categorization of them under different recommendation scenarios, which is shown in Fig.~\ref{overview_of_methods}. Specifically, the four gray boxes represent three unstudied recommendation scenarios and a recommendation scenario that is equivalent to single-domain recommendation for which we have no further method-based classification. In the following sections, we will introduce existing works under $3$ recommendation scenarios, i.e., user non-overlap $\&$ item non-overlap, user partial overlap $\&$ item non-overlap (equivalent to user non-overlap $\&$ item partial overlap), and user full overlap $\&$ item non-overlap (equivalent to user non-overlap $\&$ item full overlap), in turn. 

\begin{table*}[t]
	\caption{Method-based categorization of existing approaches for the recommendation scenario with user non-overlap and item non-overlap.}
	\label{user_item_non_overlap}
	\centering
	\setlength{\tabcolsep}{3 mm}
	\begin{tabular}{c|c|c|c|c|c}
	\toprule
	\multirow{2}{*}{\textbf{Method}}& \multirow{2}{*}{\textbf{Approach}} & \multirow{2}{*}{\textbf{Venue}} & \multirow{2}{*}{\textbf{Year}} & \multicolumn{2}{c}{\textbf{Recommendation Tasks}}\\
	\cline{5-6}
	& & & & \textbf{Dimension 1} &\textbf{Dimension 2}\\
	 \midrule
	\multirow{12}{*}{\shortstack{Extracting\\Cluster-\\Level\\Rating \\Patterns}} & CBT~\cite{DBLP:conf/ijcai/LiYX09} & IJCAI & 2009 & intra-domain & single-target\\
	 & TALMUD~\cite{DBLP:conf/cikm/MorenoSRS12} & CIKM & 2012 & intra-domain & single-target\\
	 & CLFM~\cite{DBLP:conf/pkdd/GaoLCLGG13} & PKDD & 2013 & intra-domain & multi-target\\
	 & PCLF~\cite{DBLP:conf/aaai/RenGLG15} & AAAI & 2015 & intra-domain & multi-target\\
	 & MINDTL~\cite{DBLP:conf/wsdm/HeZYY18} & WSDM & 2018 & intra-domain & single-target\\
	 & FUSE~\cite{DBLP:conf/kdd/ChenHL13} & KDD & 2013 & intra-domain & single-target\\
	 & ProbKT~\cite{DBLP:conf/iconip/ZhangWLZ18} & ICONIP. & 2018 & intra-domain & single-target\\
	 & CIT~\cite{DBLP:conf/um/ShiLH11} & Support Syst. & 2017 & intra-domain & single-target\\
	 & CrossFire~\cite{DBLP:conf/wsdm/ShuWTWL18} & WSDM & 2018 & intra-domain & single-target\\
	 & RMGM~\cite{DBLP:conf/icml/LiYX09} & ICML & 2009 & intra-domain & multi-target\\
	 & ~\cite{DBLP:conf/aistats/IwataK15} & AISTATS & 2015 & intra-domain & multi-target\\
	 & CDIE-C~\cite{DBLP:conf/wsdm/WangFGCH19} & WSDM & 2019 & intra-domain & multi-target\\
	 \midrule
	\multirow{9}{*}{\shortstack{Capturing\\Tag\\Correlations}} & UserItemTags~\cite{DBLP:conf/ecweb/EnrichBR13} & EC-Web & 2013 & intra-domain & single-target\\
	 & TagGSVD++~\cite{DBLP:conf/recsys/Fernandez-TobiasC14} & RecSys & 2014 & intra-domain & single-target\\
	 & TagCDCF~\cite{DBLP:conf/um/ShiLH11} & UMAP & 2011 & intra-domain & multi-target\\
	 & ETagiCDCF~\cite{DBLP:conf/fuzzIEEE/HaoZL16} & FUZZ-IEEE & 2016 & intra-domain & multi-target\\
	 & SCT~\cite{DBLP:conf/ijcnn/ZhangH0019} & IJCNN & 2019 & intra-domain & multi-target\\
	 & TMT~\cite{DBLP:conf/icdm/FangGLLL15} & ICDMW & 2015 & intra-domain & multi-target\\
	 & SCD~\cite{DBLP:conf/cidm/KumarKHCA14} & CIDM & 2014 & inter-domain & single-target\\
	 & GRAPH~\cite{DBLP:conf/cikm/YangHQXW15} & CIKM & 2015 & inter-domain & single-target\\
	 & CFAA~\cite{DBLP:journals/corr/abs-2202-04920} & WWW & 2022 & intra-domain & multi-target\\
	 \midrule
	\multirow{3}{*}{\shortstack{Applying\\Active\\Learning}} & MMMF$_{TL}$~\cite{DBLP:conf/aaai/ZhaoPXZLY13} & AAAI & 2013 & intra-domain & single-target\\
	 & RLMF$_{TL}$, PMF$_{TL}$~\cite{DBLP:journals/ai/ZhaoPY17} & Artif. Intell. & 2017 & intra-domain & single-target\\
	 & MultiAL~\cite{DBLP:conf/aaai/ZhangJLD016} & AAAI & 2016 & intra-domain & multi-target\\
	\bottomrule
	\end{tabular}
\end{table*}

\section{Scenario 1: User non-overlap $\&$ Item non-overlap}
\label{scenario_1}
Due to the independence and isolation of information in different domains, user sets and item sets of two domains are not overlapped, or even there are some overlapped users or items, they can not be identified and the correspondences are not available. This leads to the first recommendation scenario, where neither user sets nor item sets overlap, which was extensively studied in the early cross-domain recommendation. In particular, there are three classes of approaches for this recommendation scenario, that is, ($\romannumeral1$) \emph{extracting cluster-level rating patterns}, ($\romannumeral2$) \emph{capturing tag correlations}, and ($\romannumeral3$) \emph{applying active learning technology}. The method-based categorization of existing approaches is displayed in Table~\ref{user_item_non_overlap}.

\subsection{Extracting Cluster-Level Rating Patterns}
\label{cluster-level}

\subsubsection{The basic paradigm}
\begin{figure*}[t]
    \centering
    \includegraphics[width=\textwidth]{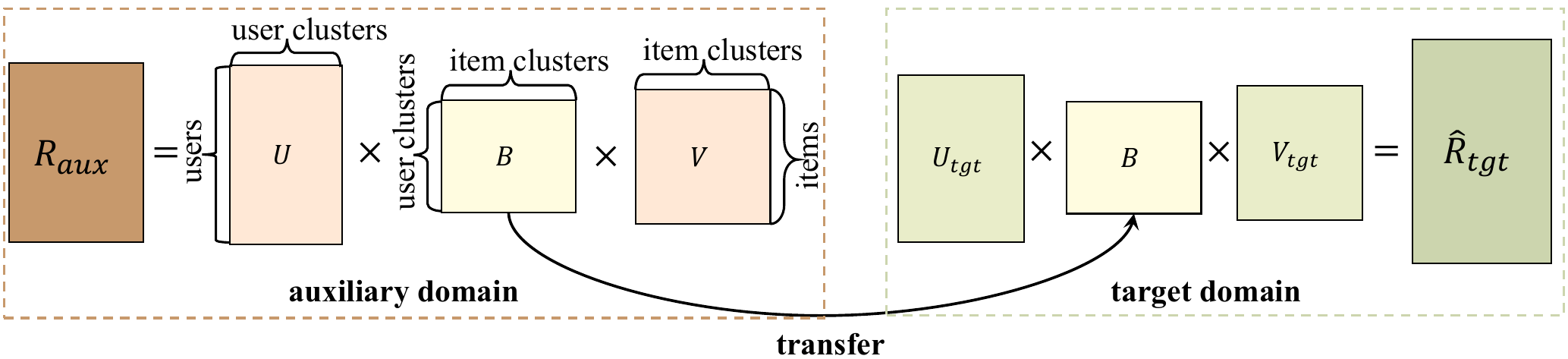}
    \caption{The schematic diagram of extracting cluster-level rating patterns.}
    \label{framework_of_cluster_level_rating_pattern}
\end{figure*}

The first class of approaches assume that the services of domains are geared towards the general population and users in different domains may have similar preferences while items may share some properties as well. Although there are not overlapping users/items  between domains, two domains may share cluster-level rating patterns. Therefore, approaches of this class aim at extracting cluster-level rating patterns from one domain and transfer them to the other domain. Fig.~\ref{framework_of_cluster_level_rating_pattern} shows the schematic diagram of this method and we describe the details in the following. 

The first step is to factorize the rating matrix of the source domain $\textbf{R}_{aux}$ into three latent matrices $\textbf{U}\in\mathbb{R}^{m^A\times k}_{+}$, $\textbf{V}\in\mathbb{R}_{+}^{n^A\times l}$, and $\textbf{S}\in\mathbb{R}^{k\times l}_{+}$. Specifically, the orthogonal nonnegative matrix tri-factorization (ONMTF) algorithm is a widely used co-clustering algorithm that is proved to be equivalent to a two-way K-means clustering algorithm. 

\begin{eqnarray}
\underset{\textbf{U}\ge0,\textbf{V}\ge0,\textbf{S}\ge0}{min}\Vert\textbf{R}_{aux}-\textbf{U}\textbf{S}\textbf{V}^T\Vert_{F}^{2}\\
s.t. \quad  \textbf{U}^{T}\textbf{U}=\textbf{I}, \quad \textbf{V}^{T}\textbf{V}=\textbf{I},  \nonumber
\end{eqnarray}

Each row of $\textbf{U}$ and $\textbf{V}$ is the cluster indicator for a user or an item. 

The next step is to construct a compact matrix $\textbf{B}$ that encodes the cluster-level rating patterns of $k$ clusters of users and $l$ clusters of items that are shared between domains. The third step is to get the cluster indicator matrices $\textbf{U}_{tgt}$ and $\textbf{V}_{tgt}$ in the target domain by transferring $\textbf{B}$ to the target domain.

\begin{eqnarray}
\label{objective_function}
\underset{\textbf{U}_{tgt}\in\{0, 1\}^{p\times k},\textbf{V}_{tgt}\in\{0, 1\}^{q\times l}}{min}\Vert[\textbf{X}_{tgt}-\textbf{U}_{tgt}\textbf{B}\textbf{V}^T_{tgt}]\circ\textbf{W}\Vert_{F}^{2},\\
s.t. \quad \quad \textbf{U}_{tgt}\textbf{1}=\textbf{1}, \quad \textbf{V}_{tgt}\textbf{1}=\textbf{1},  \nonumber
\end{eqnarray}
where $\circ$ denotes the element-wise product. Finally, the predicted ratings $\tilde{\textbf{X}}_{tgt}$ in the target domain are generated:
\begin{eqnarray}
\tilde{\textbf{X}}_{tgt} = \textbf{W}\circ\textbf{X}_{tgt}+[1-\textbf{W}]\circ[\textbf{U}_{tgt}\textbf{B}\textbf{V}_{tgt}^{T}]
\end{eqnarray}

\subsubsection{Approaches of this method} The first approach of this method is proposed by Li et al.~\cite{DBLP:conf/ijcai/LiYX09}. The generated compact matrix $\textbf{B}$ is called a ``codebook". They set the nonnegative entry of $\textbf{U}$ and $\textbf{V}$ in each row to be $1$ and the others to be $0$ getting two indicator matrices $\textbf{U}_{aux}$ and $\textbf{V}_{aux}$. Then the codebook $\textbf{B}$ is constructed by averaging all the ratings in each user-item co-cluster as an entry in the codebook as follows:

\begin{eqnarray}
\textbf{B}=[\textbf{U}^{T}_{aux}\textbf{X}_{aux}\textbf{V}_{aux}]\oslash[\textbf{U}^{T}_{aux}11^{T}\textbf{V}_{aux}] 
\end{eqnarray}

The basic paradigm was then widely borrowed, expanded, and improved by many researchers, and some variant approaches have been proposed.

Moreno et al.~\cite{DBLP:conf/cikm/MorenoSRS12} extended this paradigm by transferring knowledge from multiple source domains with varying levels of relevance and proposed an approach named TALMUD. For each source domain with rating matrix $\textbf{X}_{src_n}$, there is a codebook $\textbf{B}_n$ representing the transferred knowledge from this domain and a variable $\alpha_n$ denoting its relevant degree with the target domain. The target rating matrix is reconstructed by linearly integrating the rating patterns of all source domains, i.e., the equation~(\ref{objective_function}) is changed as follows:

\begin{eqnarray}
\underset{\textbf{U}_{n}\in\{0, 1\}^{p\times k_n},\textbf{V}_{n}\in\{0, 1\}^{q\times l_n},\alpha_n\in\textbf{R}}{min}\Vert[\textbf{X}_{tgt}-\sum_{n=1}^N\alpha_n(\textbf{U}_{n}\textbf{B}_n\textbf{V}^T_{n})]\circ\textbf{W}\Vert_{F}^{2}\\
s.t. \quad \quad \textbf{U}_{n}\textbf{1}=\textbf{1}, \quad \textbf{V}_{n}\textbf{1}=\textbf{1},  \nonumber
\end{eqnarray}

Some researchers proposed that there simultaneously existed common cluster-level rating patterns shared across domains and domain-specific cluster-level rating patterns for each domain. The final predicted rating scores should be the combination of ratings from the perspective of common rating patterns and ratings from the perspective of domain-specific rating patterns. Following this idea, Gao et al.~\cite{DBLP:conf/pkdd/GaoLCLGG13} proposed a Cluster-Level Latent Factor Model (CLFM) which partitioned the cluster-level rating patterns $\textbf{B}$ in each domain into a common part $\textbf{B}_0$ and a domain-specific part $\textbf{B}_\tau$, that is $\textbf{B}_{\tau} = [\textbf{B}_0, \textbf{B}_\tau]$. Ren et al.~\cite{DBLP:conf/aaai/RenGLG15} proposed a Probabilistic Cluster-level Latent Factor (PCLF) model that jointly learnt a common cluster-level rating matrix and a domain-specific cluster-level rating matrix. Zhang et al.~\cite{DBLP:conf/iconip/ZhangWLZ18} proposed a joint probabilistic model named ProbKT which jointly captured domain-shared group-level knowledge and domain-specific group-level knowledge. The equation~(\ref{objective_function}) of these approaches is changed as follows:

\begin{eqnarray}
\underset{\textbf{U}_{n}\in\{0, 1\}^{p\times k_n},\textbf{V}_{n}\in\{0, 1\}^{q\times l_n},\alpha_n\in\textbf{R}}{min}\sum_{\tau}\Vert[\textbf{X}_{\tau}-\textbf{U}_{\tau}[\textbf{B}_0,\textbf{B}_{\tau}]\textbf{V}^T_{\tau}]\circ\textbf{W}_{\tau}\Vert_{F}^{2}\\
s.t. \quad \quad \textbf{U}_{\tau}\textbf{1}=\textbf{1}, \quad \textbf{V}_{\tau}\textbf{1}=\textbf{1},  \nonumber
\end{eqnarray}

Chen et al.~\cite{DBLP:conf/kdd/ChenHL13} extended this paradigm by taking users' generated tags into consideration and proposed a tensor-factorization-based framework (FUSE). Instead of sharing a cluster-level rating matrix between domains, they constructed a shared three-dimensional cluster-level tensor to transfer knowledge. The matrix factorization process is, therefore, replaced by tensor factorization which maps users, items, and tags into a shared latent feature space. The clustering operation is simultaneously performed on users, items, and tags to obtain clusters of them. 

Zhang et al.~\cite{DBLP:journals/dss/ZhangWLLZ17} proposed that there are divergences between domains and directly transferring the cluster-level rating patterns of the source domain to the target domain may result in "negative transfer". Therefore, they utilized a domain adaptation technique to ensure the consistency of the transferred knowledge. He et al.~\cite{DBLP:conf/wsdm/HeZYY18} proposed that the orthogonal nonnegative matrix tri-factorization (ONMTF) algorithm adopted by existing approaches requires that the matrix be fully rated which is not easy to be satisfied. They extended the TALMUD~\cite{DBLP:conf/cikm/MorenoSRS12} approach by applying an incomplete orthogonal nonnegative matrix tri-factorization (IONMTF) algorithm which relaxed the full rating restriction on the rating matrix of ONMTF and was easier to implement on real-world datasets.

Shu et al.~\cite{DBLP:conf/wsdm/ShuWTWL18} proposed a cross-media joint friend and item recommendation framework (CrossFire) which aimed to jointly recommend items and friends to users in the target domain. In addition to tri-factorize rating matrices into user feature matrices, item feature matrices, and a shared rating feature matrix as previous works, it also tri-factorized user-user link matrices into user feature matrices and a shared interaction matrix. Besides, CrossFire also utilized item features as auxiliary information. It assumed that items in different domains share a dictionary and factorized feature matrices into the item feature matrices and the dictionary in each domain.

\subsubsection{Expand to multi-target recommendation.}
All of the approaches mentioned above are for single-target recommendation as they explicitly extract cluster-level rating patterns in the source domain and then transfer them to the target domain. Some researchers have tried to expand this method to multi-target recommendation. 

Li et al.~\cite{DBLP:conf/icml/LiYX09} proposed a rating-matrix generative model (RMGM) which integrated several sparse domains and assumed that multiple domains share common latent cluster-level patterns. Then they learned the probabilities of each user and item belonging to this shared latent structure. Iwata et al.~\cite{DBLP:conf/aistats/IwataK15} proposed that the latent factors of users/items generated by matrix factorization in each domain were from a common Gaussian distribution. The same mean vector and covariance matrix helped to align the latent factors from different domains.

Wang et al.~\cite{DBLP:conf/wsdm/WangFGCH19} focused on the cross-domain session-based recommendations and proposed a method called cross-domain item embedding method based on co-clustering (CDIE-C). As user information is not available and items differ between domains, we also categorize it to the recommendation scenario where neither user nor item overlaps. The core idea of CDIE-C is to extract cluster-level correlations by exploring cross-domain co-occurrence relations of items based on the co-clustering method. Then, both item-level sequence information within each domain and cluster-level cross-domain correlation information can be captured to generate the final cross-domain recommendation.

\subsection{Capturing Tag Correlations}
\subsubsection{The basic paradigm.}

\begin{figure*}[t]
    \centering
    \includegraphics[width=\textwidth]{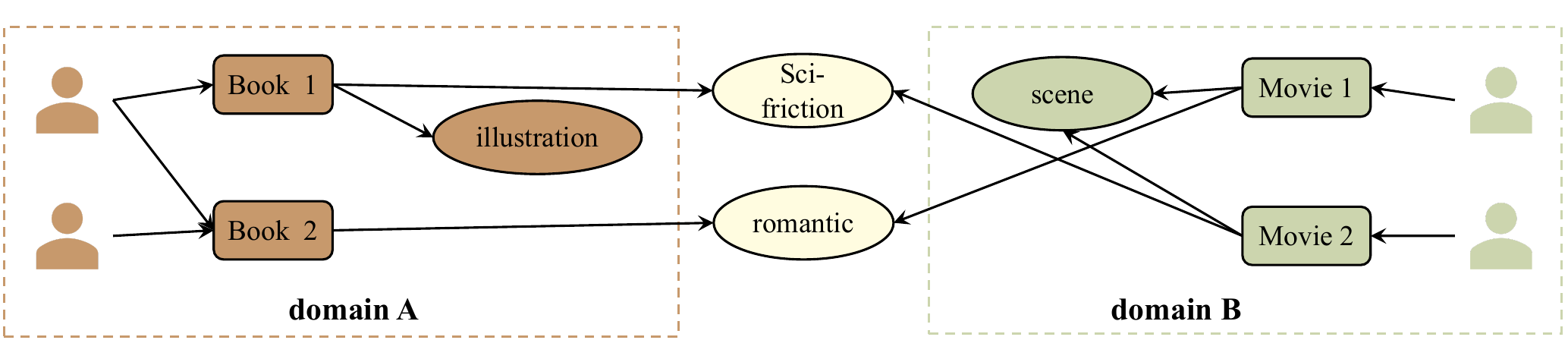}
    \caption{The schematic diagram of capturing tag correlations.}
    \label{framework_of_capturing_tag_correlations}
\end{figure*}

This method is based on the assumption that although users and items are different between domains, users may use the same tags to annotate items of interest, and items in different domains may be tagged by the same tags to encode their properties. As shown in Fig.~\ref{framework_of_capturing_tag_correlations}, domain A and domain B have different users and items. Users in domain A use tags $\{illustration, sci-friction, romantic\}$ to tag books, and users in domain B use tags $\{scene, sci-friction, romantic\}$ to tag movies, so both tags ``sci-friction'' and ``romantic'' exist in both domains. Therefore, this method turns to user-generated tags to establish the linkages between different domains. Specifically, there are two ways to capture tag correlations. On the one hand, tags can be used to simultaneously enhance the profiles of users and items. On the other hand, matrices of similarities between users, items, and tags can be generated based on tags and then be used as constraints to learn better user and item representations.

\subsubsection{Approaches of this method}
For enhancing the profiles of users and items, Enrich et al.~\cite{DBLP:conf/ecweb/EnrichBR13} first proposed to utilize tags as implicit user feedback to enhance item factors. They proposed a tag-based cross-domain collaborative filtering approach based on the SVD++~\cite{DBLP:conf/kdd/Koren08} algorithm with three different adaptations being explored. Fernandez-Tobias et al.~\cite{DBLP:conf/recsys/Fernandez-TobiasC14} claimed that the approach proposed in~\cite{DBLP:conf/ecweb/EnrichBR13} did not fully exploit users' preferences expressed in their tags assigned to items. They further proposed an approach named TagGSVD++ based on GSVD++~\cite{DBLP:conf/sac/Manzato13} which simultaneously enriched users' profiles with tags they used and extended items' profiles with tags they received.

For constructing similarity matrices between users and items, Shi et al.~\cite{DBLP:conf/um/ShiLH11} proposed a tag-induced cross-domain collaborative filtering (TagCDCF) algorithm that exploited shared tags to construct cross-domain user-to-user similarity matrix $S^{(U)}$ and item-to-item similarity matrix $S^{(V)}$. The tag-induced similarity matrices were then incorporated into the matrix factorization process as additional constraints by forcing similar users/items to have closer representations. The objective function is generally defined as follows:

\begin{eqnarray}
F(U^{(1)},V^{(1)},U^{(2)},V^{(2)})&=&\frac{1}{2}\sum_{k=1}^{2}I^{R^{(k)}}(R^{(k)}-U^{(k)T}V^{(k)})^2 \\ \nonumber
&+&\frac{\alpha}{2}I^{S^{(V)}}(S^{(V)}-V^{(1)T}V^{(2)})^2\\ \nonumber
&+&\frac{\alpha}{2}I^{S^{(U)}}(S^{(U)}-U^{(1)T}U^{(2)})^2. \\ \nonumber
\end{eqnarray}

Many researchers have proposed improved approaches based on this idea. Hao et al.~\cite{DBLP:conf/fuzzIEEE/HaoZL16} proposed that the number of shared tags between domains was limited and it was a waste to discard domain-dependent tags. They proposed an Enhanced Tag-induced Cross-Domain Collaborative Filtering (ETagiCDCF) algorithm to explore domain-dependent tags. It first grouped domain-dependent tags into clusters based on their co-occurrences with shared tags. The cross-domain user-to-user and item-to-item similarities were computed on these tag clusters. Zhang et al.~\cite{DBLP:conf/ijcnn/ZhangH0019} proposed that although there were a limited number of shared tags between domains, two non-identical tags in two domains might be semantically related. They proposed a cross-domain recommendation approach with semantic correlations in tagging systems (SCT). It utilized word2vec to analyze all tags and generated continuous vectors to capture contextual information and semantic similarities between tags. Both the intra-domain similarities and inter-domain similarities between users and items can be computed based on tags' semantic similarities.

Instead of constructing the similarity matrices between users and items, Fang et al.~\cite{DBLP:conf/icdm/FangGLLL15} proposed to construct a tag co-occurrence matrix that captured the interrelatedness among tags. Then the rating matrix in each domain was tri-factorized into three parts, that is, the tag co-occurrence matrix, the user latent factors for tags, and the item latent factors for tags. The factorization process was optimized by forcing the generated co-occurrence matrix to be as close to the constructed co-occurrence matrix as possible. 
Liu et al.~\cite{DBLP:journals/corr/abs-2202-04920} recently extended this method by treating user/item features as bridges to link domains instead of utilizing tags. The rating prediction module jointly combines users'/items' IDs, historical ratings, and reviews to generate their general embeddings and perform recommendations based on the embeddings. The embedding attribution alignment module treats each dimension of an embedding as a certain kind of attribution. Both vertical and horizontal alignments are equipped to reduce the domain discrepancy and transfer consistent useful knowledge across the source and target domains.

\subsubsection{Expand to Inter-domain recommendation}

Kumar et al.~\cite{DBLP:conf/cidm/KumarKHCA14} proposed that the vocabularies of different domains may not match explicitly, but through the use of ontologies, it may be possible to derive semantic relationships between words of distinct domains. They proposed a semantic-clustering-based cross-domain (SCD) recommendation algorithm which first performed semantic clustering to obtain clusters of semantically equivalent words. SCD then got item-based topic distributions and user-based topic distributions. The user profiles were then mapped to the target domain and used to perform inter-domain recommendations.

Yang et al.~\cite{DBLP:conf/cikm/YangHQXW15} had a similar proposal that different vocabularies were used by different domains so tags should be correlated on semantic level rather than lexical level. Therefore, they utilized online encyclopedias to achieve the semantic matching of tags and built a multi-partite graph to represent the similarities of objects in different domains. Finally, the similarity between a user and an item was measured based on the graph propagation.

\subsection{Applying Active Learning}
\subsubsection{The basic paradigm.}
This method involves a certain amount of human effort within a fixed budget. For example, at first, there were no explicit entity (i.e., user or item) correspondences between different domains, some users/items are sometimes overlapped. It is expensive or time-consuming to recognize all user correspondences, but the partial mappings of a small number of users or items can be identified within a fixed budget. In addition, more ratings can be obtained through human efforts, which alleviates the problem of data sparsity. 

\subsubsection{Approaches of this method.}
Zhao et al.~\cite{DBLP:conf/aaai/ZhaoPXZLY13} proposed a margin-based active learning approach that selected the entities in the target domain with low prediction certainty. They first applied a maximum-margin matrix factorization (MMMF) in the target domain to get the original user and item latent factors. They then queried the correspondences of entities with low prediction certainty in the source domain and an extended MMMF was proposed to transfer knowledge from the source domain to refine the user/item latent factors in the target domain. This method was later generalized to a unified framework. Two variants based on regularized low-rank matrix factorization (i.e., RLMF$_{TL}$) and probabilistic matrix factorization (i.e., PMF$_{TL}$) were proposed in their later works~\cite{DBLP:journals/ai/ZhaoPY17}. 

Zhang et al.~\cite{DBLP:conf/aaai/ZhangJLD016} incorporated active learning with the previous proposed RMGM~\cite{DBLP:conf/icml/LiYX09} approach. It first generated the original user and item latent factors as well as shared cluster-level rating patterns using the RMGM algorithm. It added unrated user-item pairs of all domains into a pool and the proposed algorithm iteratively selected the most informative items from the pool to ask for users' ratings. The selecting criterion measured the global generalization errors which jointly considered domain-specific errors and domain-independent errors. The new rated user-item pairs will be added to the training set and help to re-train the RMGM algorithm.

\subsection{Discussion}
In this section, we have classified existing approaches for the recommendation with user non-overlap and item non-overlap into three kinds of methods. Now, we would like to make a comparison between them and elaborate about when and why each method performs better, and what are their limitations.

For the \textit{Extracting Cluster-Level Rating Patterns} method, knowledge transfer between domains is at cluster-level (or group-level) instead of at user-level. Such knowledge transfer is coarse-grained, which will limit the recommendation performance of the approaches. However, this method is universal when there are no overlapping entities (either users or items) between domains. 
For the \textit{Capturing Tag Correlations} method, domains are explicitly associated with tags, and approaches of this method generally perform well. At the same time, the ideas proposed by this method can be further used to build the connections between domains utilizing comments and external knowledge. However, such approaches are no longer applicable when additional information is not available.
For the \textit{Applying Active Learning} method, it is to invest a number of human resources to change the scenario where users and items do not overlap into the scenario where users or items overlap partially, so the recommendation performance can be improved. However, these approaches require a budget, and the performance is affected by the quantity and quality of the identified entity pairs.

\section{Scenario 2: User partial overlap $\&$ Item non-overlap}
\label{scenario_2}

\begin{table*}[t]
	\caption{Method-based categorization of existing approaches for the recommendation scenario with user partial overlap and item non-overlap}
	\label{user_partially_item_non}
	\centering
	\setlength{\tabcolsep}{2.5 mm}
	\begin{tabular}{c|c|c|c|c|c}
	\toprule
	\multirow{2}{*}{\textbf{Method}}& \multirow{2}{*}{\textbf{Approach}} & \multirow{2}{*}{\textbf{Venue}} & \multirow{2}{*}{\textbf{Year}} & \multicolumn{2}{c}{\textbf{Recommendation Tasks}}\\
	\cline{5-6}
	& & & & \textbf{Dimension 1} &\textbf{Dimension 2}\\
	 \midrule
	 \multirow{5}{*}{\shortstack{Collective\\Matrix\\Factorization}}& XPT{\scriptsize{RANS}} \cite{DBLP:conf/aaai/JiangCYXY16} & AAAI & 2016 & intra-domain & multi-target\\ 
	 &JCSL~\cite{DBLP:conf/pkdd/RafailidisC16} & PKDD & 2016 & intra-domain & multi-target\\
	 & CDCR~\cite{DBLP:conf/cikm/RafailidisC17} & CIKM & 2017 & intra-domain & multi-target\\
	 & KerKT~\cite{DBLP:journals/tnn/ZhangLWZ19} & - & 2019 & intra-domain & multi-target\\
	 & MPF~\cite{DBLP:conf/sigir/YangYYLC17} & SIGIR & 2017 & intra-domain & multi-target\\	
	\midrule
	 \multirow{5}{*}{\shortstack{Representation\\Combination\\of Overlapping\\Users}}& DTCDR~\cite{DBLP:conf/cikm/ZhuC0LZ19} & CIKM & 2019 & intra-domain & multi-target\\	 
	 & GA-DTCDR~\cite{DBLP:conf/ijcai/ZhuWCLZ20} & IJCAI & 2020 & intra-domain & multi-target\\
	 & GA-DTCDR-P~\cite{DBLP:journals/corr/abs-2108-07976} & TKDE &  & intra-domain & multi-target\\ 
	 & ~\cite{DBLP:conf/mm/PereraZ17} & Multimedia & 2017 & intra-domain & single-target\\
	 & ~\cite{DBLP:conf/aaai/PereraZ20} & AAAI & 2020 & intra-domain & single-target\\
	\midrule
	 \multirow{12}{*}{\shortstack{Embedding\\and\\Mapping}}& EMCDR~\cite{DBLP:conf/ijcai/ManSJC17} & IJCAI & 2017 & inter-domain & single-target\\	 
	 & CDLFM~\cite{DBLP:conf/dasfaa/WangPWYFH18} & DASFAA & 2018 & inter-domain & single-target\\
	 & RC-DFM~\cite{DBLP:conf/aaai/FuPWXL19} & AAAI & 2019 & inter-domain & single-target\\
	 & DCDIR~\cite{DBLP:conf/sigir/BiSYWWX20a} & SIGIR & 2020 & inter-domain & single-target\\
	 & HCDIR~\cite{DBLP:conf/sigir/BiSYWWX20} & SIGIR & 2020 & inter-domain & single-target\\
	 & DCDCSR~\cite{DBLP:conf/ijcai/ZhuWCLOW18} & IJCAI & 2018 & inter-domain & single-target\\
	 & SSCDR~\cite{DBLP:conf/cikm/KangHLY19} & CIKM & 2019 & inter-domain & single-target\\
	 & TMCDR~\cite{DBLP:conf/sigir/ZhuGZXXZL021} & SIGIR & 2021 & inter-domain & single-target\\
	 & PTUPCDR~\cite{DBLP:conf/wsdm/ZhuTLZXZLH22} & WSDM & 2022 & inter-domain & single-target\\
	 &LACDR~\cite{DBLP:conf/cikm/WangZZWZH21} & CIKM & 2021 & inter-domain & single-target\\
	 & CGN~\cite{DBLP:conf/ijcai/ZhangLHMCLT20} & IJCAI & 2020 & intra-domain & multi-target\\
	 &DML~\cite{DBLP:journals/corr/abs-2104-08490} & TKDE & 2021 & intra-domain & multi-target\\
	\midrule
	\multirow{4}{*}{\shortstack{Graph Neural\\ Network-based\\Approaches}} & HeroGRAPH~\cite{DBLP:conf/recsys/CuiWZZ20} & IJCAI & 2017 & intra-domain & multi-target\\
	& NSCR~\cite{DBLP:conf/sigir/Wang0NC17} & SIGIR & 2017 & inter-domain & single-target\\
	& ReCDR~\cite{DBLP:conf/cikm/XuXCZ21} & CIKM & 2021 & intra-domain & multi-target\\ 
	& ECHCDR~\cite{DBLP:conf/dasfaa/LiPWXYH20} & DASFAA & 2020 & intra/inter-domain & multi-target\\
	\midrule
	\multirow{2}{*}{\shortstack{Capturing\\Aspect Correlations}}&\multirow{2}{*}{CATN~\cite{DBLP:conf/sigir/ZhaoLXDS20}} & \multirow{2}{*}{SIGIR} & \multirow{2}{*}{2020} & \multirow{2}{*}{inter-domain} & \multirow{2}{*}{multi-target}\\
	& & & & & \\
	\bottomrule
	\end{tabular}
\end{table*}

For this kind of recommendation scenario, some users have interactions in both domains while others are only in a specific domain. It is symmetrical to the scenario with item partial overlap and user non-overlap. Approaches discussed in this section can be applied to these two recommendation scenarios. In particular, these approaches can be categorized into five classes, that is, ($\romannumeral1$) \emph{collective matrix factorization}, ($\romannumeral2$) \emph{representation combination of overlapping users}, ($\romannumeral3$) \emph{embedding and mapping}, ($\romannumeral5$) \emph{graph neural network-based approaches}, and ($\romannumeral4$) \emph{capturing aspect correlations}. A brief summary of method-based categorization of existing approaches for this recommendation scenario is displayed in Table~\ref{user_partially_item_non}.

\subsection{Collective Matrix Factorization}
\label{collective_matrix_factorization}

\begin{figure*}[t]
    \centering
    \includegraphics[width=.65\textwidth]{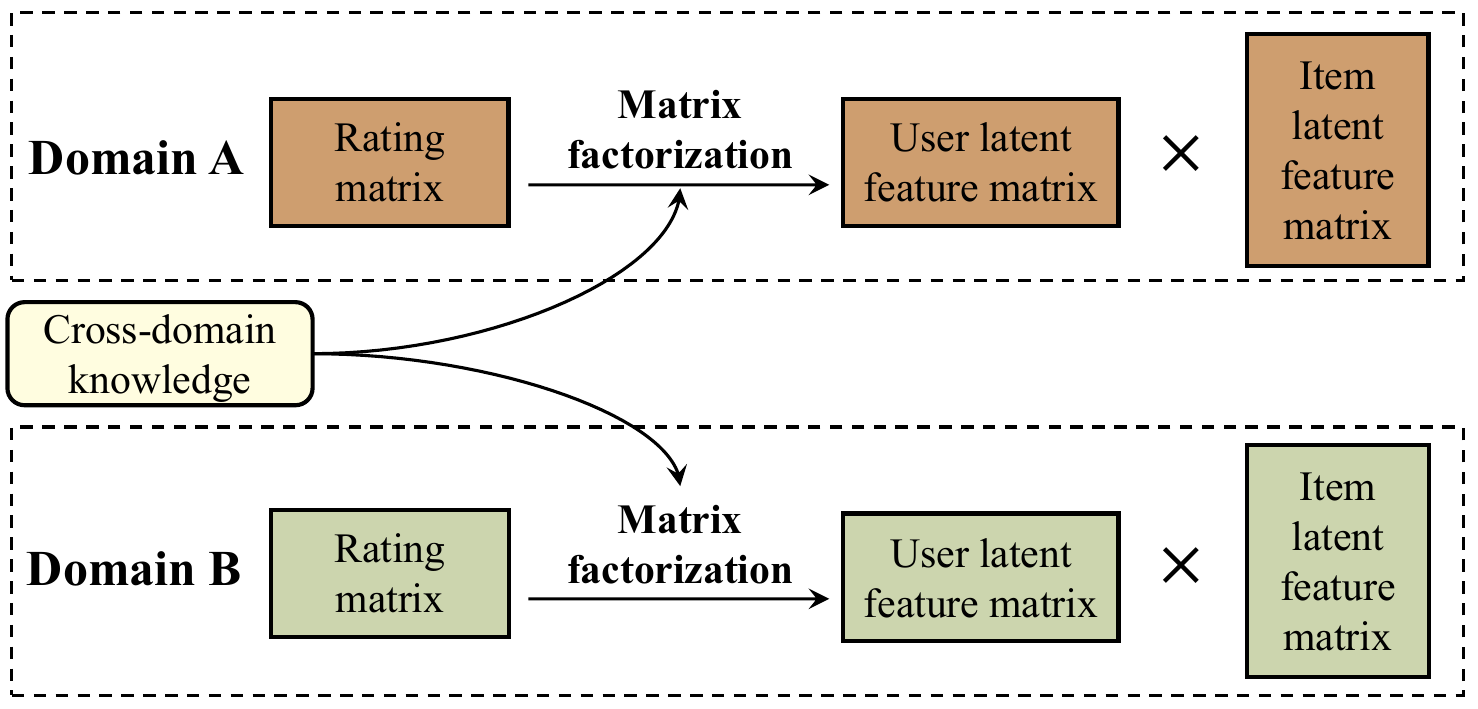}
    \caption{The schematic diagram of collective matrix factorization.}
    \label{framework_of_collective_matrix_factorization}
\end{figure*}

\subsubsection{The basic paradigm.}
Collective matrix factorization is a direct extension of the general matrix factorization method for the cross-domain recommendation problem. Fig.~\ref{framework_of_collective_matrix_factorization} shows the schematic diagram of collective matrix factorization. Similar to general matrix factorization, the rating matrix in each domain is factorized into a user latent feature matrix and an item latent feature matrix. The difference is that cross-domain knowledge is utilized to constrain the matrix factorization process within each domain.

\subsubsection{Approaches of this method.}
Jiang et al.~\cite{DBLP:conf/aaai/JiangCYXY16} proposed a semi-supervised transfer learning approach called XPT{\scriptsize{RANS}} in which they argued that the similarities between overlapped users were consistent across different domains. They first performed nonnegative matrix factorization on two domains. Then the user-based similarities were added as constraints to the matrix factorization process when learning the representations of users and items. Rafailidis et al.~\cite{DBLP:conf/pkdd/RafailidisC16} proposed a joint cross-domain user clustering and similarity learning recommendation algorithm (JCSL) in which they jointly considered cluster-based and user-based cross-domain similarities. The resulting similarity matrices acted as social regularization terms during the matrix factorization process. Rafailidis et al.~\cite{DBLP:conf/cikm/RafailidisC17} further proposed a cross-domain recommendation approach with collaborative ranking (CDCR) which focused on the ranking performance when generating the recommendation. In this model, they learned the cross-domain user latent matrix to capture correlations of users in two domains and incorporated it as a constraint into the learning process of user/item latent factors. 

Zhang et al. extended their previous study~\cite{DBLP:journals/dss/ZhangWLLZ17} to the scenario of user partial overlap and proposed a cross-domain recommender system with kernel-induced knowledge transfer (KerKT)~\cite{DBLP:journals/tnn/ZhangLWZ19}. The overlapping users were utilized to train the domain adaptation function to ensure the consistency of the transferred knowledge. Kernel-induced completion was conducted to measure the user similarities which were integrated into the matrix factorization process as constraints. Yang et al.~\cite{DBLP:conf/sigir/YangYYLC17} proposed a generative model of Multi-site Probabilistic Factorization (MPF) the basic idea of which was to model cross-site user preferences and site-specific user preferences simultaneously. For multiple-site users, their latent feature vectors $u_i^{(s)}$ in site $s$ consist of a common part $\bar{u}_i$ and a site-specific part $\vartriangle u_i^{(s)}$, i.e., $u_i^{(s)}=\bar{u}_i+\vartriangle u_i^{(s)}$. For an exclusive user in site $s$, $u_i^{(s)}=\bar{u}_i+0$. In the generative process, each site had a different prior for the site-specific part of users' latent feature vectors.

\subsection{Representation Combination of Overlapping Users}

\subsubsection{The basic paradigm} 
\begin{figure*}[t]
    \centering
    \includegraphics[width=\textwidth]{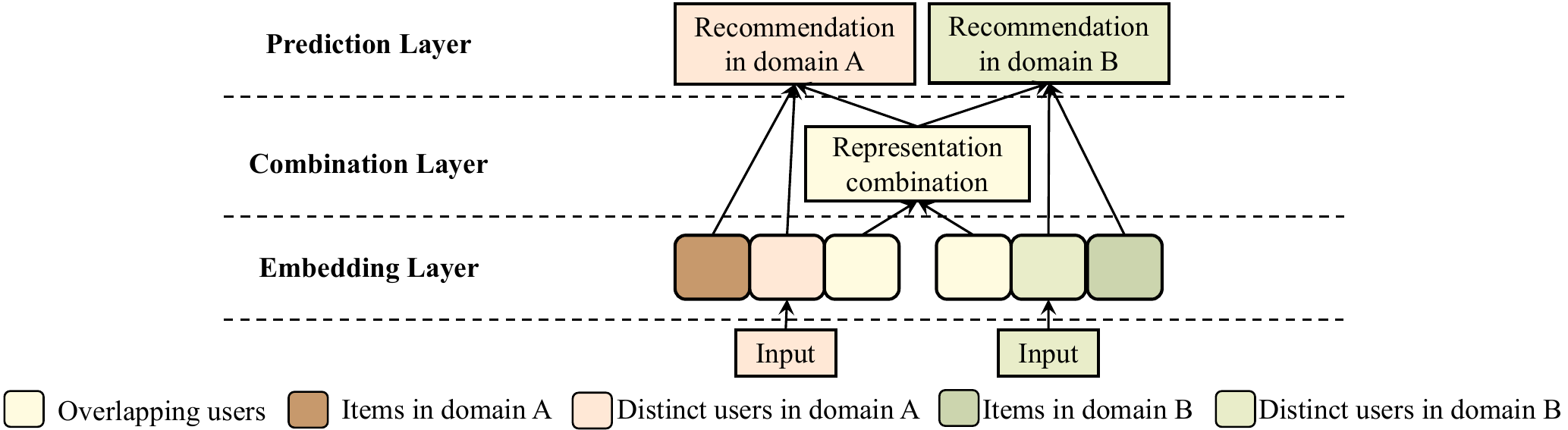}
    \caption{The schematic diagram of representation combination of overlapping users.}
\label{framework_of_representation_combination_of_overlapping_user}
\end{figure*}

Fig.~\ref{framework_of_representation_combination_of_overlapping_user} shows the schematic paradigm of this method. We can see, there are typically three layers. The \emph{embedding layer} generates embeddings for users and items in each domain. In the \emph{combination layer}, the embeddings of overlapping users from both domains are combined to generate the unified embeddings for overlapping users. Finally, the \emph{prediction layer} takes both the embeddings of distinct users and overlapping users to train the recommendation model on each domain separately. 

\subsubsection{Approaches of this method}
Zhu et al.~\cite{DBLP:conf/cikm/ZhuC0LZ19} first proposed this paradigm for dual-target cross-domain recommendation (DTCDR) the core idea of which was to share the knowledge of overlapping users across domains. Specifically, in the embedding layer, it generated embeddings for users and items in each domain from both rating and content information. In the combination layer, the embeddings of overlapping users were combined by three different combination operations, i.e., concatenation, max-pooling, and average-pooling. Zhu et al.~\cite{DBLP:conf/ijcai/ZhuWCLZ20} then proposed a Graphical and Attentional framework for Dual-Target Cross-Domain Recommendation (GA-DTCDR). In the embedding layer, it applied a graph embedding technique (i.e., Node2vec) to generate more representative user and item embeddings in each domain. In the combination layer, it employed an element-wise attention mechanism to more effectively combine the representations of overlapping users.
Zhu et al. later extended their proposed framework to support three different recommendation scenarios~\cite{DBLP:journals/corr/abs-2108-07976}, that is, dual-target CDR, multi-target CDR, and cross-domain recommendation and cross-system recommendation (CDR+CSR). For different scenarios, personalized training strategies are adopted.

Some approaches of this method focused on capturing the dynamic nature of user preferences. Perera et al. proposed to utilize time-stamped, cross-network information for both new (i.e., distinct users) and existing (i.e., overlapping users) user recommendation~\cite{DBLP:conf/mm/PereraZ17}. In the embedding layer, they generated users' topical distribution by topic modeling in each domain. Before the combination, two transfer functions map user preferences from the topical space to the target network user space. In the combination layer, the representations of overlapping users are obtained by fusing user preferences from both domains. Later, they further proposed a time-aware unified cross-network solution~\cite{DBLP:conf/aaai/PereraZ20} which, in the embedding layer, modeled user preferences under short, long and long short term levels in each domain. Then, in the combination layer, the three-level preference representations in both domains of overlapping users (referred to as existing users in their paper) were integrated to obtain users' final representations. The distinct users' (referred to as new users) representations were directly generated by fusing the three representations from the source domain.

\subsection{Embedding and Mapping}

\subsubsection{The basic paradigm}
\begin{figure*}[t]
    \centering
    \includegraphics[width=\textwidth]{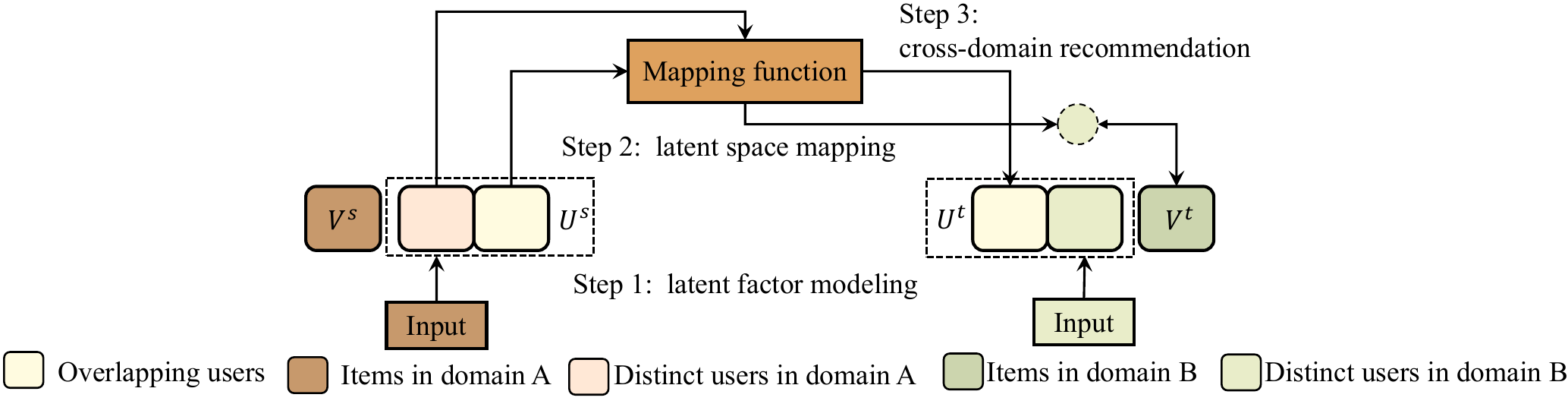}
    \caption{The schematic diagram of embedding and mapping.}
    \label{framework_of_embedding_and_mapping}
\end{figure*}

 This is an inter-domain recommendation method in which one domain is treated as the source domain and the other as the target domain.
Fig.~\ref{framework_of_embedding_and_mapping} shows the schematic diagram of this method. There are three main steps, i.e., \emph{latent factor modeling}, \emph{latent space mapping} and \emph{cross domain recommendation}. During the latent factor modeling process, the aim is to generate user and item latent factors $\{U^s, V^s, U^t, V^t\}$ in each domain. During the latent space mapping, the aim is to train a mapping function $f_U$. The objective of $f_U$ is to establish the relationships between the latent space of domains:

\begin{eqnarray}
\underset{\theta}{min}\sum_{u_i\in \mathcal{U}}L(f_U(U_i^s;\theta), U_i^t),
\end{eqnarray}

During the cross-domain recommendation process, for a user who only has a latent factor in the source domain, it generates the user's latent factor in the target domain:

\begin{eqnarray}
\hat{U}_i^t=f_U(U_i^s;\theta),
\end{eqnarray}

With $\hat{U}_i^t$, the recommendation in the target domain can be performed to this user.

\subsubsection{Approaches of this method}
The embedding and mapping framework for cross-domain recommendation (EMCDR) was first proposed by Man et al.~\cite{DBLP:conf/ijcai/ManSJC17} that aimed at the \emph{inter-domain recommendation} problem. For latent factor modeling, EMCDR, respectively, applied Matrix Factorization (MF) and Bayesian Personalized Ranking (BPR) to generate user and item latent factors. For latent space mapping, it utilized both a linear function and a nonlinear function based on Multi-Layer Perceptron (MLP) to act as the mapping function. The objective is to approximate the latent factor $U_i^s$ of overlapping users in the source domain mapped by $f_U$ with their corresponding latent factors $U_i^t$ in the target domain. 

This idea was widely improved by many researchers and the improvement is mainly in two aspects, one is the latent factor modeling process, the other is the latent space mapping process. 

For latent factor modeling, Wang et al.~\cite{DBLP:conf/dasfaa/WangPWYFH18} proposed a Cross-Domain Latent Feature Mapping (CDLFM) model. It first defined three similarity measurements on users' rating behaviors. During latent factor modeling, the similarity values were embedded into the matrix factorization process as constraints. Fu et al.~\cite{DBLP:conf/aaai/FuPWXL19} proposed a Review and Content-based Deep Fusion Model (RC-DFM). It extended stacked denoising autoencoders to effectively fuse review text and item contents with the rating matrix to generate user and item representations with more semantic information. Bi et al.~\cite{DBLP:conf/sigir/BiSYWWX20a,DBLP:conf/sigir/BiSYWWX20} proposed to construct a heterogeneous information network and took into consideration the interaction sequence information to learn effective user/item representations in each domain. The proposed approaches are proved to be effective in the cross-domain insurance recommendation.

For latent space mapping, Zhu et al.~\cite{DBLP:conf/ijcai/ZhuWCLOW18} proposed a Deep framework for both Cross-Domain and Cross-System Recommendation (DCDCSR) which took into account rating sparsity degrees of individual users/items to generate benchmark factor matrices. The mapping function was trained to map the latent factor matrices of users and items to fit the benchmark factor matrices.
Kang et al.~\cite{DBLP:conf/cikm/KangHLY19} proposed that in existing EMCDR-based approaches, the training of mapping functions only used overlapping users, so their performance was sensitive to the number of overlapping users. After an in-depth analysis of the Amazon dataset, they proposed that in real-world datasets, the number of overlapping users was always small, which limited the performance of existing approaches. Therefore, they proposed a Semi-Supervised framework for Cross-Domain Recommendation (SSCDR) to utilize both the overlapping users and source-domain items to train the mapping function. 
Zhu et al. also proposed that the mapping function is biased to the limited overlapping users, which degrades the generalization ability of the model. The proposed transfer-meta framework for CDR (TMCDR)~\cite{DBLP:conf/sigir/ZhuGZXXZL021} can replace the training procedure of most existing EMCDR-based methods and has good generalization ability.
Zhu et al. later proposed that the complex relationships between user preferences of the source and target domains vary from user to user, which is hard to be captured by a single mapping function. Their proposed framework named Personalized Transfer of User Preferences for Cross-domain Recommendation (PTUPCDR)~\cite{DBLP:conf/wsdm/ZhuTLZXZLH22} learns a meta-network to personalize a mapping function for each user.
Wang et al. pointed out a similar problem and proposed an approach namely Low-dimensional Alignment for Cross-Domain Recommendation (LACDR)~\cite{DBLP:conf/cikm/WangZZWZH21}. Instead of learning a mapping function utilizing overlapping users, LACDR first trains an encoder and a decoder in both the source domain and the target domain utilizing both overlapping users and non-overlapping users. For a cold-start user, the representation in the target domain is obtained by first generating her domain-invariant factor using the encoder of the source domain. Then, the domain-invariant factor is mapped to the latent factor in the target domain using the decoder in the target domain.

\subsubsection{Expand to intra-domain recommendation}
The approaches described above are for inter-domain recommendation, as they all learn a mapping function between different domains and map the users' representation in the source domain to the target domain. Some works have applied this embedding and mapping framework to the intra-domain recommendation.

Zhang et al. proposed that users' interests and states may vary over time and it was important to quickly capture these changes for timely and accurate recommendations. Therefore, they first divided the sequence of users' interacted items into itemsets based on timestamps of interactions to denote users' interests during a period. Instead of capturing the mapping relationships between user representations in different domains like previous EMCDR-based approaches, the proposed cycle generation network (CGN)~\cite{DBLP:conf/ijcai/ZhangLHMCLT20} learned a user's personalized dual-direction mapping function between the representation of her interaction itemsets in different domains at the same temporal period.
Li et al. proposed a Dual Metric Learning (DML)~\cite{DBLP:journals/corr/abs-2104-08490} model which supplemented the dual learning mechanism with the metric learning approach. Through a pre-trained autoencoder network, DML first generate user/item embeddings $W_{u_A}, W_{u_B}, W_{i_A}, W_{i_B}$ in each domain. Instead of learning a mapping function from the source domain to the target domain as most EMCDR-based methods, DML aims to learn a transitional metric learning mapping $X$ which is restricted to be an orthogonal mapping. Embeddings of overlapping users $W_{{ou}_A}$ and $W_{{ou}_B}$ are utilized to learn this metric learning mapping. The objective is to minimize the distance between the mapped user embedding $XW_{{ou}_A}$ and the target user embeddings $W_{{ou}_B}$ for the same overlapping users, the dual form of which is to minimize the distance between $X^TW_{{ou}_B}$ and $X_{{ou}_A}$. Finally, the recommendation is performed on $XW_{u_A}$ and $W_{i_A}$ in domain A and $X^TW_{u_B}$ and $W_{i_B}$.

\subsection{Graph Neural Network-based Approaches}

\begin{figure*}[t]
    \centering
    \includegraphics[width=\textwidth]{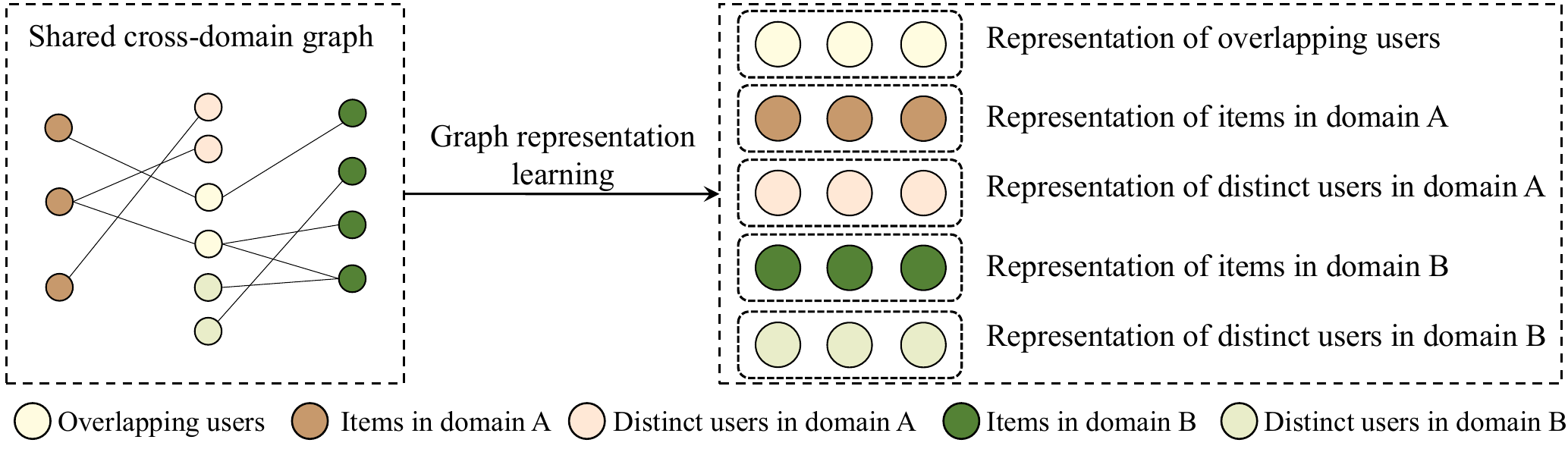}
    \caption{The schematic diagram of graph neural network-based approaches.}
    \label{framework_of_graph_neural_network}
\end{figure*}

\subsubsection{The basic paradigm.}
The basic paradigm of this method is building shared graphs to represent the relationships among users, items, attributions, and other factors (i.e., nodes in the graphs), and learn a representation for each node through graph representation learning. The learned representations are proved to be able to capture the high-order and non-linear dependencies between nodes and can be used for subsequent recommendations. Since constructing shared graphs among domains can integrate the information of different domains, through graph representation learning, the representations of nodes from different domains are embedded in the same latent space and, therefore, the cross-domain information can be transferred. Fig.~\ref{framework_of_graph_neural_network} shows the schematic diagram of this method.

\subsubsection{Approaches of this method.}
Wang et al.~\cite{DBLP:conf/sigir/Wang0NC17} first applied graph neural networks to the cross-domain social recommendation and proposed a Neural Social Collaborative Ranking (NSCR) approach. It aims to recommend items in an information-oriented domain to users in a social-oriented domain, which can be seen as an inter-domain recommendation. NSCR first utilizes an attributed-aware deep collaborative filtering model in the information-oriented domain to learn user-item interactions. Then embeddings of overlapping users are directly transferred to the social-oriented domain. The cross-domain knowledge is transferred by a multi-layer graph convolution network propagating representations of overlapping users to non-overlapping users. Finally, the embeddings of non-overlapping users in the social-oriented domain and embeddings of items in the information-oriented domain are used for recommendation.

Cui et al. proposed that, in previous works, user behaviors are processed within each domain which is an indirect way to incorporate cross-domain information. They proposed a heterogeneous graph framework (HeroGRAPH)~\cite{DBLP:conf/recsys/CuiWZZ20} that collected user behaviors from all domains to devise a shared graph to directly model users' cross-domain behaviors. Information within each domain was utilized to conduct within-domain modeling and graph convolution operations with recurrent attention on the shared graph is applied to conduct cross-domain modeling. The within-domain representations and cross-domain representations of users and items are combined to perform the final recommendation. 
In a similar way, Xu et al. proposed Relation Expansion based Cross-Domain Recommendation (ReCDR)~\cite{DBLP:conf/cikm/XuXCZ21}. The main difference lies in the way of constructing the shared graph. They first applied a graph embedding model (i.e., Node2Vec) to generate pre-trained node embeddings for all nodes in both domains. Nodes with higher similarities will be connected to devise the shared graph.
Li et al.~\cite{DBLP:conf/dasfaa/LiPWXYH20} further proposed an embedding content and heterogeneous network (ECHCDR) which creatively incorporated an adversarial learning algorithm. It first utilized Doc2vec to generate content representations of users/items and they are concatenated with adjacency representations to act as initial representations of users/items. It then simultaneously train a generator and a discriminator to learn suitable representations of users and items. Finally, both intra-domain and inter-domain recommendation can be performed based on the inner product of the learned representations.

\subsection{Capturing Aspect Correlations}

\subsubsection{The basic paradigm.}
Apart from the previously introduced embedding and mapping-based approaches, capturing aspect correlations is another method for inter-domain recommendation. It assumes users' preferences are multi-faceted and aims at modeling fine-grained semantic aspects and exploring their mutual relationships across domains. The score prediction can be performed by matching the aspect features of a user in one domain and an item in the other domain.

\subsubsection{Approaches of this method.}
Zhao et al.~\cite{DBLP:conf/sigir/ZhaoLXDS20} proposed a cross-domain recommendation framework via aspect transfer network (CATN) to capture users' multi-faceted and fine-grained preferences. It first represents a user by a user document that contains all reviews written by this user, and an item by an item document that contains all reviews it receives. It then generates abstract aspect features for each user and each item from their documents. Aspect features of overlapping users were utilized to identify the global cross-domain aspect correlations. The inter-domain recommendation can be performed by utilizing the user's review document in the source domain and the item's review document in the target domain, and vice versa. Specifically, the rating prediction is obtained by aggregating the semantic matching between two aspects in the aspect features of a user and an item.

\subsection{Discussion}
In this section, we have classified existing approaches for the recommendation with user partial overlap and item non-overlap. Now, we would like to make a comparison between them and elaborate about when and why each method performs better, and what are their limitations.

For the \textit{Collective Matrix Factorization} method, cross-domain knowledge is incorporated into the matrix factorization process as constraints. The performance of these approaches is often inferior to that of deep learning-based methods.
For the \textit{Representation Combination of Overlapping Users} method, it fully integrates the representations (also referred to as knowledge) of overlapping users in different domains. These approaches generally perform well when the number of overlapping users is sufficient. However, there are no special designs for non-overlapping users and the representations of them are generated in the same way as in single-domain recommendations. The performance of these approaches will deteriorate when non-overlapping users make up the majority of the dataset.
For the \textit{Graph Neural Network-based} method, it fuses the information of both domains by constructing a shared graph and propagating cross-domain knowledge through graph learning technologies, which achieves the optimal performance in many cases. However, due to the high demand for computing resources, the scalability of these approaches may be limited.
For the \textit{Capturing Aspect Correlations} method, it can exploit users' multi-faceted and fine-grained aspect-level preferences by utilizing content information (e.g., user profiles and comments). Such approaches transfer knowledge across domains in a more granular manner and show better performance. However, this method is not applicable when content information is not available.  
Finally, for the \textit{Embedding and Mapping} method, it is originally a customized framework for cross-domain cold-start recommendation (i.e., inter-domain recommendation). When extended to the intra-domain recommendation, it also achieved satisfactory performance. The performance of these approaches depends on the design of mapping functions and whether sufficient samples are available to train mapping functions.

\section{Scenario 3: User full overlap $\&$ Item non-overlap}
\label{scenario_3}

\begin{table*}[t]
	\caption{Method-based categorization of existing approaches for the recommendation scenario with user fully overlap and item non-overlap}
	\label{user_fully_item_non}
	\centering
	\setlength{\tabcolsep}{3.9 mm}
	\begin{tabular}{c|c|c|c|c|c}
	\toprule
	\multirow{2}{*}{\textbf{Method}}& \multirow{2}{*}{\textbf{Approach}} & \multirow{2}{*}{\textbf{Venue}} & \multirow{2}{*}{\textbf{Year}} & \multicolumn{2}{c}{\textbf{Recommendation Tasks}}\\
	\cline{5-6}
	& & & & \textbf{Dimension 1} &\textbf{Dimension 2}\\
	 \midrule
	 \multirow{4}{*}{\shortstack{Collective\\Matrix\\Factorization}}& CMF~\cite{DBLP:conf/kdd/SinghG08} & KDD & 2008 & intra-domain & multi-target\\
	 & SoRec~\cite{DBLP:conf/cikm/MaYLK08a} & CIKM & 2008 & intra-domain & multi-target\\
	 & CTR+RBF~\cite{DBLP:conf/ijcai/XinLLHWG15} & IJCAI & 2015 & intra-domain & multi-target\\
	 & LSCD~\cite{DBLP:conf/dasfaa/ZhaoHWH18} & DASFAA & 2018 & intra-domain & multi-target\\
	\cline{1-6}
	 \multirow{3}{*}{\shortstack{Tensor\\Factorization}}& CDTF~\cite{DBLP:conf/www/HuCXCGZ13} & WWW & 2013 & intra-domain & multi-target\\
	 & HST~\cite{DBLP:conf/icml/LiuHW15} & ICML & 2015 & intra-domain & multi-target\\
	 & RB-JTF~\cite{DBLP:conf/dasfaa/SongPWFHY17} & DASFAA & 2017 & intra-domain & multi-target\\
	 \cline{1-6}
	 \multirow{2}{*}{\shortstack{Factorization\\Machines}}& FM-MCMC~\cite{DBLP:conf/ecir/LoniSLH14} & ECIR & 2014 & intra-domain & single-target\\
	 & CoFM~\cite{DBLP:conf/aaai/LiDL19} & AAAI & 2019 & intra-domain & single-target\\
	 \cline{1-6}
	 \multirow{9}{*}{\shortstack{Deep\\Sharing \\ User\\ Representations}}& MVDNN~\cite{DBLP:conf/www/ElkahkySH15} & WWW & 2015 & intra-domain & multi-target\\
	 & CCCFNet~\cite{DBLP:conf/www/LianZXS17} & WWW & 2017 & intra-domain & multi-target\\
	 & GCBAN~\cite{DBLP:conf/icdm/HeLZLNH18} & ICDM & 2018 & intra-domain & multi-target\\
	 & PPGN~\cite{DBLP:conf/cikm/ZhaoLF19} & CIKM & 2019 & intra-domain & multi-target\\
	 & DA-GCN~\cite{DBLP:conf/ijcai/0008T0ZNY21} & IJCAI & 2021 & intra-domain & multi-target \\
	 & DeepAPF~\cite{DBLP:conf/ijcai/YanCGLJ19} & IJCAI & 2019 & intra-domain & multi-target\\
	 & EATNN~\cite{DBLP:conf/sigir/ChenZWMLLM19} & SIGIR & 2019 & intra-domain & multi-target\\
	 & MSDCR~\cite{DBLP:conf/wsdm/ZhaoYY22} & WSDM & 2022 & intra-domain & multi-target \\
	 & \cite{DBLP:conf/kdd/LiYMZLGDW21} & KDD & 2021 & intra-domain & multi-target \\
	\cline{1-6}
	 \multirow{5}{*}{\shortstack{Deep\\Dual\\Knowledge\\Transfer}}& CoNet~\cite{DBLP:conf/cikm/HuZY18} & CIKM & 2018 & intra-domain & multi-target\\
	 & ACDN~\cite{DBLP:conf/www/LiuZZLSX0X20} & WWW & 2020 & intra-domain & multi-target\\
	 & DDTCDR~\cite{DBLP:conf/wsdm/0008T20} & WSDM & 2020 & intra-domain & multi-target\\
	  & $\pi$-Net~\cite{DBLP:conf/sigir/MaRLCMR19} & SIGIR & 2019 & intra-domain & multi-target\\
  	 & BiTGCF~\cite{DBLP:conf/cikm/LiuLLP20} & CIKM & 2020 & intra-domain & multi-target\\
	 \cline{1-6}
	 \multirow{5}{*}{\shortstack{Deep Integration of \\ Source Domain \\ Information}}& MTNet~\cite{hu2018mtnet} & KDD & 2018 & intra-domain & single-target\\
	 & NATR~\cite{DBLP:conf/www/GaoCFZ00J19} & WWW & 2019 & intra-domain & single-target\\
	 & MF$\_$S~\cite{DBLP:conf/ijcai/MaZWLLM18} & IJCAI & 2018 & intra-domain & single-target\\
	 & ~\cite{DBLP:conf/ijcai/PereraZ18} & IJCAI & 2018 & intra-domain & single-target\\
	 & PriCDR~\cite{DBLP:journals/corr/abs-2202-04893} & WWW & 2022 & intra-domain & single-target \\
	\bottomrule
	\end{tabular}
\end{table*}

This category of recommendation refers to the scenario in which all users have interactions in all domains while items are disjoint among domains. The users are sometimes called multi-homed users which are widely studied in recent years.
This recommendation scenario is symmetrical to the scenario with item full overlap and user non-overlap. The approaches discussed in this section can be applied to these two recommendation scenarios. In particular, approaches can be divided into six classes, that is, ($\romannumeral1$) \emph{collective matrix factorization}, ($\romannumeral2$) \emph{tensor factorization}, and ($\romannumeral3$) \emph{factorization machines}, ($\romannumeral4$)\emph{deep sharing user representations}, ($\romannumeral5$)\emph{deep dual knowledge transfer}, and ($\romannumeral6$) \emph{deep integration of source domain information}. Specifically, the first three classes of approaches are generally in shallow structures while the last three classes are deep learning-based approaches. 
A brief summary of method-based categorization of existing approaches for this recommendation scenario is displayed in Table~\ref{user_fully_item_non}.

\subsection{Collective Matrix Factorization}

\subsubsection{The basic paradigm.}
As described in section~\ref{collective_matrix_factorization}, collective matrix factorization is a direct extension of the traditional matrix factorization method for cross-domain recommendation problem. The difference is that cross-domain knowledge can be utilized to constrain the matrix factorization process within each domain. To some extent, the approaches discussed in section~\ref{collective_matrix_factorization} are applicable to this recommendation scenario. We omit the schematic diagram as it is the same as Fig.~\ref{framework_of_embedding_and_mapping}. In the following, we discuss approaches that are specially proposed for this recommendation scenario. 

\subsubsection{Approaches of this method.}
Singh et al.~\cite{DBLP:conf/kdd/SinghG08} first proposed the collective matrix factorization (CMF) model which collectively factorized rating matrices $R^A, R^B$ of two domains into user representations $U_A,U_B$ and item representations $V_A,V_B$. The constraint was that the representations of users are shared across different domains which means $U_A=U_B$. Ma et al.~\cite{DBLP:conf/cikm/MaYLK08a} proposed a probabilistic matrix factorization approach for social recommendation (SoRec). The rating matrix was factorized into a user feature matrix $U_A$ and an item feature matrix $V_A$ while the social network matrix was factorized into a user feature matrix $U_B$ and a factor feature matrix $Z_B$. Similar to CMF, the cross-domain constraint was that the user feature matrix was shared during the process, i.e., $U_A=U_B$.

However, both of the above two approaches force the same user to have exactly the same representations in different domains, that is, they assume that users' characteristics and preferences are consistent, which ignores users' domain-specific characteristics. To overcome this shortcoming, Xin et al.~\cite{DBLP:conf/ijcai/XinLLHWG15} proposed a relaxed restriction by putting forward the nonlinear mapping relationships between user representations. They proposed to train two nonlinear mapping functions $f(\cdot), g(\cdot)$ to make $f(U_A)=U_B$ and $g(U_B)=U_A$. Zhao et al.~\cite{DBLP:conf/dasfaa/ZhaoHWH18,DBLP:journals/ijon/HuangZWHC19} proposed a low-rank and sparse cross-domain (LSCD) recommendation approach in which the user latent feature matrices $U_A, U_B$ were divided into two parts: domain-shared feature matrix $U^S$ and domain-specific feature matrix $H_A, H_B$. The user representation is the sum of these two parts, i.e., $U_A=U^S+H_A, U_B=U^S+H_B$. 

\subsection{Tensor Factorization}

\begin{figure*}[t]
    \centering
    \includegraphics[width=\textwidth]{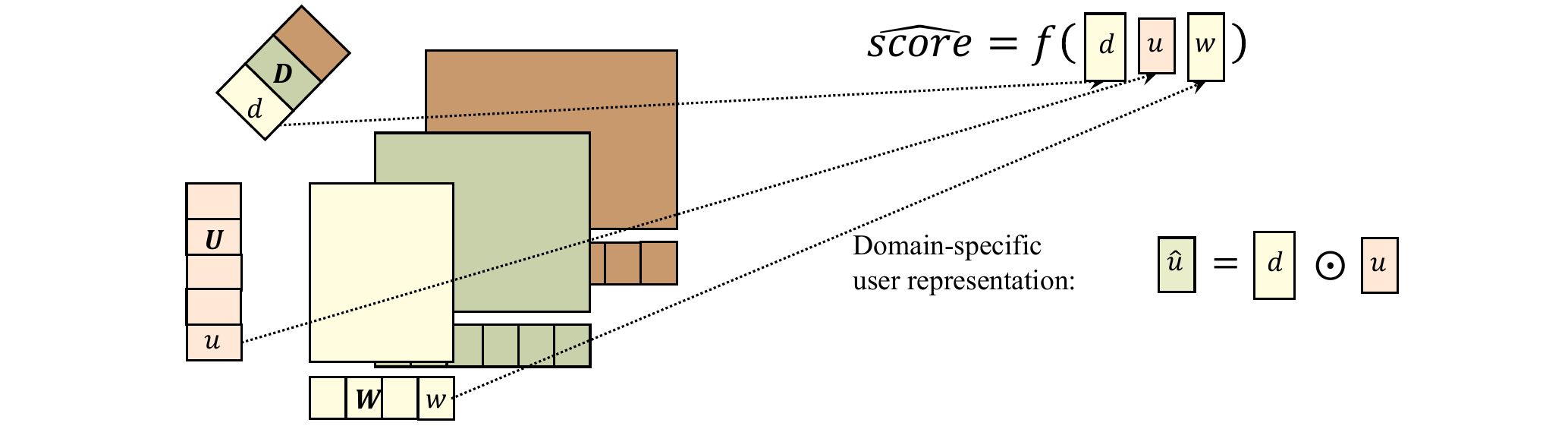}
    \caption{The schematic diagram of tensor factorization-based approaches.}
    \label{framework_of_tensor_factorization}
\end{figure*}

\subsubsection{The basic paradigm.}
In essence, tensor factorization is the higher-order generalization of matrix factorization. When only user factors and item factors are considered, the interaction information within each domain can be represented by a matrix. However, when domain factors, time factors, and aspect factors are further considered, the information within each domain will change from a matrix to a tensor. The core idea is to factorize the tensor into user representations, item representations, and other factor representations (i.e., domain, time, and aspect). By multiplying the user representations and domain representations, the domain-specific representations can be obtained. Similarly, time-specific representations and aspect-specific representations can be obtained. The final prediction $\hat{score}$ is derived from the product of multiple related representations. CP model (canonical decomposition/parallel factor analysis (PARAFAC)) is the most widely used tensor factorization algorithm. Fig.~\ref{framework_of_tensor_factorization} shows the schematic diagram of tensor factorization-based approaches.

\subsubsection{Approaches of this method.}
Hu et al.~\cite{DBLP:conf/www/HuCXCGZ13} proposed a Cross-Domain Triadic Factorization (CDTF) model which took into consideration the full triadic relation user-item-domain to reveal user preferences on items within various domains in depth. Except for a user-factor matrix $U$ shared among domains and an exclusive item-factor matrix $V_k$ for each domain, CDTF also generated a domain-factor matrix $D$ to express the traits of each domain. Then the recommendations are performed based on the results of triadic interactions among user, item, and domain factors. Song et al. proposed to exploit the aspect factors extracted from the review text to improve the performance of cross-domain recommendation. They extracted fine-grained user preferences in aspect-level and concern degrees toward different aspects of items as two tensors. A review-based joint tensor factorization (RB-JTF) approach~\cite{DBLP:conf/dasfaa/SongPWFHY17} was proposed and tensor factorization was applied simultaneously in each domain. The rating matrices were factorized into user, item, and aspect latent factors while knowledge transfer was realized by sharing user latent factors and transferring aspect latent factors. 

Liu et al. proposed that previous studies only identified and transferred the linearly correlated knowledge between domains. They proposed a new knowledge transfer technique, called the hyper-structure transfer (HST)~\cite{DBLP:conf/icml/LiuHW15}, that captured the non-linear correlations of knowledge between domains. Compared with previous approaches that directly share the cluster-level rating patterns between domains (see section~\ref{cluster-level}), HST first generated a more complex structure $\mathcal{H}$ and required the transferred rating pattern matrices in each domain to be projections of $\mathcal{H}$. To get $\mathcal{H}$, canonical polyadic decomposition of tensors is applied.

\subsection{Factorization Machines}
\subsubsection{The basic paradigm.}
Factorization Machines (FM) is a machine learning algorithm based on matrix factorization that aims to solve the problem of feature combination in large-scale sparse data. The advantages of FM fit well with the problem scenario of recommendation systems and FM has been proved to be one of the effective algorithms with verified effects. Formula (\ref{factorization_machines}) shows a second-order factorization machine model which can be easily extended to multi-order models. 

\begin{eqnarray}
\label{factorization_machines}
y=\omega_0+\sum^{n}_{i=1}\omega_i x_i + \sum^{n-1}_{i=1}\sum^{n}_{j=i+1}\omega_{ij}x_i y_j,
\end{eqnarray}
where $n$ represents the dimension of the feature vector. $x_i$ denotes the $i$-th feature, and $w_{ij}$ is the combination parameter representing the importance of the combination feature.

FM has been already used for single-domain recommendation, but for cross-domain recommendation, it is necessary to extend FM to allow them to incorporate user interaction patterns from different domains.

\subsubsection{Approaches of this method.}
Loni et al.~\cite{DBLP:conf/ecir/LoniSLH14} proposed an extension of FM named FM-MCMC that incorporated user domain-specific interaction patterns from the source domain to expand feature vectors of the target domain. The expanded feature vectors served as input to the general FM model. Specifically, a domain-dependent real-valued function was defined to control the amount of knowledge that was transferred from the source domain. Li et al.~\cite{DBLP:conf/aaai/LiDL19} designed coupled factorization machines (CoFM) and proposed that the coupled fields of coupled datasets contained shared characteristics as well as domain-specific uniqueness. CoFM, therefore, allowed the latent vectors of the coupled field between two domains to have a shared part and a domain-specific part.

\subsection{Deep Sharing User Representations}

\subsubsection{The basic paradigm} 
\begin{figure*}[t]
    \centering
    \includegraphics[width=\textwidth]{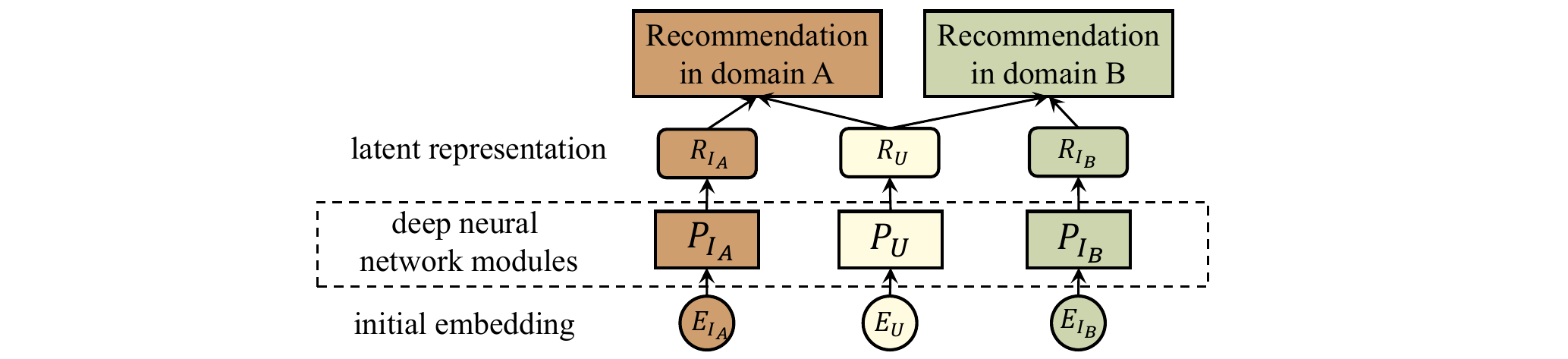}
    \caption{The schematic diagram of deep sharing user representations.}
    \label{framework_of_deep_sharing}
\end{figure*}

The schematic diagram of this class of approaches is shown in Fig.~\ref{framework_of_deep_sharing}. The core idea is to first generate users' initial embeddings $E_U$, items' initial embeddings $E_{I_A}$ and $E_{I_B}$ in domain $A$ and domain $B$, respectively. Then these initial embeddings, respectively, enter into three separate deep neural network modules $P_U, P_{I_A}, P_{I_B}$ to learn the latent representations $R_U, R_{I_A}, R_{I_B}$ of users and items. The module $P_U$ that deals with users is shared between two domains to achieve deep sharing of user representations.

\subsubsection{Approaches of this method}
Elkahky et al.~\cite{DBLP:conf/www/ElkahkySH15} proposed a Multi-View Deep Neural Network (MVDNN) in which the initial embeddings were generated from users' features and items' attributes. User embeddings were input of one view while item embeddings were input of another two separate views. All the views shared the same structure of multi-layer perception to generate representations of users and items. Two domains shared the same user view, thus achieving the goal of deep sharing of user representations. Following this approach, Lian et al. and He et al., respectively, proposed a cross-domain content-boosted collaborative filtering neural network (CCCFNet)~\cite{DBLP:conf/www/LianZXS17} and a general cross-domain framework via a bayesian neural network (GCBAN)~\cite{DBLP:conf/icdm/HeLZLNH18}. The core idea of these two approaches was to incorporate both collaborative filtering and content-based factors into account when generated initial embeddings of users and items in each view. Thus, the initial embeddings consist of two parts: one part generated from features and the other part generated from one-hot encoding. The structure of each view is the same as MVDNN~\cite{DBLP:conf/www/ElkahkySH15}.

Zhao et al.~\cite{DBLP:conf/cikm/ZhaoLF19} proposed a Preference Propagation GraphNet (PPGN) which constructed a cross-domain preference matrix to model the interactions of different domains as a whole. The high-order user preferences were propagated and integrated through multiple graph convolution and propagation layers. Compared with previous works, the multi-layer perception was replaced by multiple graph convolutional layers.
Guo et al. further proposed another graph-based solution, namely Domain-Aware Graph Convolutional Network (DA-GCN)~\cite{DBLP:conf/ijcai/0008T0ZNY21}, for the shared-account cross-domain sequential recommendation. They constructed a cross-domain sequence graph to link different domains and designed two attention mechanisms during the message passing process.

Some researchers improved previous works by jointly considering users' domain-shared representations and domain-specific representations while only domain-shared representations are shared across domains. Yan et al.~\cite{DBLP:conf/ijcai/YanCGLJ19} proposed a model of Deep Attentive Probabilistic Factorization (DeepAPF) in which the user embeddings were initialized into three parts: $P_U^S$, $P_U^A$, and $P_U^B$. $P_U^S$ was the domain-shared representations capturing cross-domain commonality of user interests and $P_U^A, P_U^B$ were the domain-specific representations capturing site-peculiarity of user interests. Attention mechanisms were utilized to fuse these two kinds of user representations to generate users' final representations, i.e., $\hat{P}_U^A=Attention(P_U^S,P_U^A), \hat{P}_U^B=Attention(P_U^S,P_U^B)$. 
Similarly, Chen et al.~\cite{DBLP:conf/sigir/ChenZWMLLM19} proposed an efficient adaptive transfer neural network (EATNN) for the social-aware recommendation which jointly performed item recommendation and friend recommendation. It initialized user embeddings into a domain-shared part $\textbf{P}_U^C$, an item domain-specific part $\textbf{P}_U^I$, and a social domain-specific part $\textbf{P}_U^S$. Two attention-based kernels were utilized to automatically estimate the difference of mutual influences between the item domain and the social domain to incorporate the shared and domain-specific representations in each domain.
Zhao et al.~\cite{DBLP:conf/wsdm/ZhaoYY22} proposed a Multi-Sparse-Domain Collaborative Recommendation (MSDCR) method which learns user domain-specific and domain-invariant preference in a more fine-grained manner (i.e., aspect-level). A User's domain-specific preference in a domain is enhanced by transferring the user's complementary aspect preferences in other domains. Domain adaption is applied to capture users' domain-invariant aspect preferences that are shared across domains.
Li et al.~\cite{DBLP:conf/kdd/LiYMZLGDW21} focused on mitigating domain biases when transferring user information across domains. The debiasing parameters contain the propensity score which solves the bias in single domain by re-weighting each transaction in the observed data. Users' general preference $\textbf{v}_u$ is generated by the nonlinear combination of the domain-specific preference $\textbf{v}_u^d$ and the propensity score.  

\subsection{Deep Dual Knowledge Transfer}

\subsubsection{The basic paradigm} 
\begin{figure*}[t]
    \centering
    \includegraphics[width=.4\textwidth]{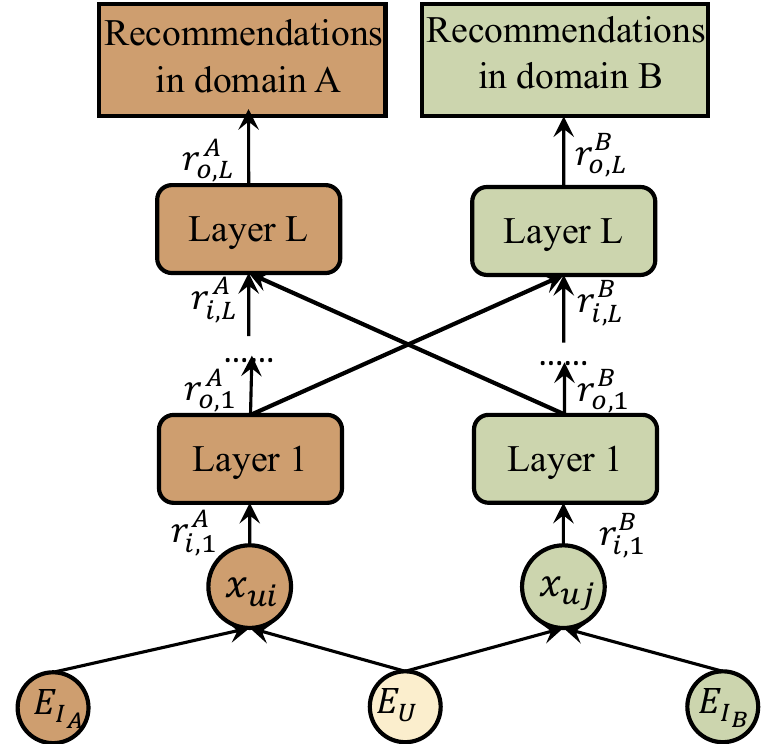}
    \caption{The schematic diagram of deep dual knowledge transfer.}
    \label{framework_of_deep_transfer}
\end{figure*}

Approaches of this class have symmetrical model structures, that is to say, each domain has the same structure and the model structures are in deep structure with multiple hidden layers. As the structures are symmetrical, these methods can simultaneously perform recommendations on multiple domains and are, therefore, for the multi-target recommendation. Fig.~\ref{framework_of_deep_transfer} shows the schematic diagram of these approaches. Let $S^A, S^B$ denote the deep neural network structure with $L$ layers in the domain $A$ and domain $B$. $r^A_{i,l}$ and $r^{A}_{o,l}$ denote the input and output of the $l$-th layer in domain $A$ while $r^B_{i,l}$ and $r^B_{o,l}$ for domain $B$. The deep dual knowledge transfer is realized by fusing the outputs of the previous layer in this domain and the other domain as the input of the next layer, which can be denoted as $r^A_{i,l+1}=Fuze^A(r^A_{o,l}, r^B_{o,l})$ and $r^B_{i,l+1}=Fuze^B(r^A_{o,l}, r^B_{o,l})$. The inputs of the first layer, i.e., $r^A_{i,1}$ and $r^B_{i,1}$, are the original embeddings of users and items. The outputs of the last layer, i.e., $r^A_{o,L}$ and $r^B_{o,L}$, are used for recommendation.

\subsubsection{Approaches of this methods}
Hu et al.~\cite{DBLP:conf/cikm/HuZY18} first proposed a collaborative cross-network (CoNet) that introduced cross-connections from one base network to another. The inputs of domain $A$ are the one-hot embeddings $x_u$ and $x_i$ of a user $u$ and an item $i$ while the inputs of domain $B$ are the one-hot embeddings $x_u$ and $x_j$ of the same user $u$ and another item $j$. The one-hot embeddings in each domain are concatenated as $x_{u,i}$ and $x_{u,j}$ acting as the input of the first layer, that is, $r^A_{i,1} = x_{u,i} = [x_u, x_i]$ and $r^B_{i,1}=x_{u,j}=[x_u,x_j]$. The deep dual knowledge transfer is achieved in multi-layer feedforward networks by adding dual connections. The formulas are as follows:

\begin{equation}
\label{deep_dual_transfer}
r^A_{i,l+1}=\sigma(W^A_lr^A_{o,l}+H_lr^B_{o,l}), \quad  r^B_{i,l+1}=\sigma(W^B_lr^B_{o,l}+H_lr^A_{o,l})
\end{equation}
where $W^A_l$ and $W^B_l$ are domain-specific parameters while $H_l$ are domain-shared parameters.

Following CoNet, Liu et al.~\cite{DBLP:conf/www/LiuZZLSX0X20} proposed to capture users' aesthetic preferences when transferring knowledge between different domains. They proposed a deep aesthetic cross-domain network (ACDN) which utilized a pre-trained deep convolutional neural network to extract aesthetic features $x_i^a$ and $x_j^a$ from item images in each domain. The aesthetic features together with one-hot embedding features $x_u$, $x_i$, $x_j$ acted as the input of the first layer, that is, 
$r^A_{i,1} = x_{u, i}=[x_u, x_i, x_i^a]$ and $r^B_{i,1}=x_{u, j} = [x_u, x_j, x_j^a]$. The deep dual knowledge transfer is realized in the same way as CoNet which is shown in equation \ref{deep_dual_transfer}.

Li et al.~\cite{DBLP:conf/wsdm/0008T20} proposed a Deep Dual Transfer Cross-Domain Recommendation (DDTCDR) model and, compared with CoNet, the improvements mainly lay in three aspects. First, DDTCDR used pre-trained autoencoders which encoded user and item features to generate feature representations as the model input. Second, it learned a latent orthogonal mapping function for transferring user preferences between domains. Third, it jointly modeled users' within-domain preferences $r_{within}$ and cross-domain preferences $r_{cross}$ to model user preferences and user-item interactions in each domain.

Ma et al. applied this method to the shared-account cross-domain sequential recommendation and proposed a parallel information-sharing network ($\pi$-Net)~\cite{DBLP:conf/sigir/MaRLCMR19}. It first utilized separate RNNs in each domain to encode user behavior sequences into a sequence representation $h_{A_i}$ and $h_{B_j}$. A shared account filter unit and a cross-domain transfer unit are designed to generate transformed representation $h^{CTU}_{{(A\rightarrow B)}_j}$ based on $h_{A_i}$ (or $h^{CTU}_{{(B\rightarrow A)}_i}$ based on $h_{B_j}$). The recommendation is performed on the concatenation representation of $h_{A_i}$ and $h^{CTU}_{{(B\rightarrow A)}_i}$ (or $h_{B_j}$ and $h^{CTU}_{{(A\rightarrow B)}_j}$). 

Liu et al.~\cite{DBLP:conf/cikm/LiuLLP20} further incorporated graph neural network into this method and proposed a bi-direction transfer learning approach by using graph collaborative filtering network as the base model (BiTGCF). It constructed a user-item bipartite graph in each domain and the multi-layer feedforward networks in previous works were replaced by multi-layer graph convolutional networks. Features of both users and items were propagated by graph convolution operations. The deep dual knowledge transfer was performed between two graph convolutional networks. Similar to previous works, the user representations of the $l+1$-th layer were a combination of the output of the previous layers in both domains. 

\subsection{Deep Integration of Source Domain Information}

\subsubsection{The basic paradigm}
\begin{figure*}[t]
    \centering
    \includegraphics[width=.45\textwidth]{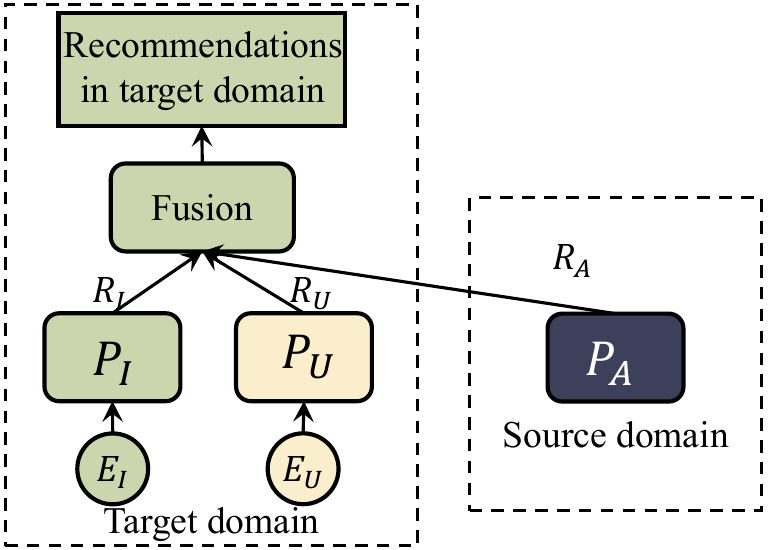}
    \caption{The schematic diagram of deep integration of source domain information.}
    \label{framework_of_deep_integration}
\end{figure*}

Different from the above two classes of approaches that are in symmetrical model structures and aim at the multi-target recommendation, this class of approaches are for the single-target recommendation. As shown in Fig.~\ref{framework_of_deep_integration}, they treat one domain as the target domain to play a dominant role and the other domains as source domains. The information $R_A$ from the source domains is incorporated into the target domain as additional auxiliary information.

\subsubsection{Approaches of this method}

Hu et al.~\cite{hu2018mtnet} proposed an approach based on neural networks named MTNet in which a memory network was utilized to generate high-level representations $\textbf{W}_o$ of the context text (i.e., product reviews) in the target domain and a transfer network was utilized to generate representations $\textbf{W}_c$ of source domain knowledge. Together with representations $\textbf{W}_z$ of the user-item interaction, these three kinds of representations were combined to perform the final recommendation. Ma et al. included users' content information in social media into rating prediction and proposed a social media content enriched matrix factorization (MF$\_$S)~\cite{DBLP:conf/ijcai/MaZWLLM18}. It first adopted several different embedding learning algorithms (i.e., word2vector, stacked denoising autoencoder) to generate user embeddings $\vec{D}_u$ from the content information. It modified users' latent vector as $\vec{U}_u=[\vec{D}_u,\vec{B}_u]$ where $\vec{B}_u$ is randomly initialized. It then applied matrix factorization to train user and item embeddings by keeping $\vec{D}_u$ unchanged and only updating $\vec{B}_u$. Perera et al.~\cite{DBLP:conf/ijcai/PereraZ18} focused on capturing the dynamic natures of user preferences and providing timely recommendations in the target domain. It first divided user interactions in source domains into several sets according to timestamps of his interactions in the target domain. After topical modeling and sum pooling, each set of interactions in source domains during a time interval was encoded as a high-level latent representation. An improved LSTM with attention mechanisms and time-aware gates were utilized to encode the sequence of the representations and generated the final recommendations.

Some works have proposed to focus on privacy protection. Gao et al. designed a model named Neural Attentive-Transfer Recommendation (NATR)~\cite{DBLP:conf/www/GaoCFZ00J19,DBLP:journals/tkdd/GaoLFCZHJ22} which shares item embeddings between domains rather than user embeddings to reduce the risk of leaking user privacy. It first looked up the embeddings in the source domain of the overlapping items that a user has interacted with within the target domain. An item-level attention unit was designed to aggregate the item embeddings generating an additional user embedding $s_u$. A domain-level attention unit was designed to fuse $s_u$ with the user embedding $p_u$ in the target domain to generate a unified user embedding $z_u$. The interaction probability was predicted by the inner product of $z_u$ and an item embedding $q_i$. 
Chen et al.  proposed a two-stage based privacy-preserving CDR framework (PriCDR)~\cite{DBLP:journals/corr/abs-2202-04893}. In stage one, two methods are proposed for the source domain to publish rating matrix differential privately to the target domain. In stage two, a deep auto-encoder and a deep neural network are utilized to model the published source rating matrix and target rating matrix, respectively, in the target domain. Embeddings of the same users in both domains are aligned.

\subsection{Discussion}
In this section, we have classified existing approaches for the recommendation with user fully overlap and item non-overlap into six kinds of methods. Among them, the first three (i.e., Collective Matrix Factorization, Tensor Factorization, and Factorization Machines) are traditional machine learning-based methods, which are in shallow structures. The last three (i.e., Deep Sharing User Representations, Deep Dual Knowledge Transfer, and Deep Integration of Source Domain Information) are deep learning-based methods. Extensive experiments have shown that methods based on deep learning generally outperform those based on traditional machine learning. This is because shallow structures are insufficient to capture complex nonlinear user-item interactions while deep neural networks endow models with better modeling abilities. 

For the three deep learning-based methods, the \textit{Deep Sharing User Representations} method utilize a shared structure to generate the representations of overlapped entities (i.e., users). This method only focuses on the similarities between domains and ignores the differences. Some approaches (i.e., DeepAPF and EATNN) overcome this problem by separating user modeling into a domain-specific part and a domain-shared part. However, the cross-domain knowledge transfer is only realized through the shared user representations, which may limit the performance of the model.
The \textit{Deep Dual Knowledge Transfer} method realizes the bidirectional deep transfer of knowledge between domains. However, as approaches of this method indiscriminately transfer knowledge from one domain to the other, irrelevant factors are easily introduced as noise, which degrades the model performance. Therefore, these approaches generally show satisfying performance when there are large similarities between domains and the data sparsity of the two domains is relatively close.
The \textit{Deep Integration of Source Domain Information} method treats the target domain as a dominant role and all other domains as additional auxiliary information to improve the recommendation of the target domain. The structure and parameters of the model are customized for the target domain, so it achieves good recommendation performance in the target domain. The limitation of these approaches is that they do not take full advantage of the complementary knowledge between domains. To make recommendations on the source domain, it is necessary to retrain the model by exchanging the roles of the source domain and the target domain.



\section{Datasets for Cross-Domain Recommendation}
\label{dataset}

In this section, we introduce datasets used in cross-domain recommendation. Specifically, we first introduce three datasets with multiple domains, namely, Amazon, Douban, and Epinions. Then we introduce datasets with a single domain such as movie, book, music, social interaction, point-of-interest, and others. Table~\ref{Datasets summary} shows the results of our classification of datasets and the usage of these datasets by existing works. In the following, we detail the source and composition of each dataset and explain how each dataset can be applied to cross-domain recommendation tasks.

\subsection{Datasets with Multiple Domains}

Datasets that contain information from multiple domains are the most widely used datasets in cross-domain recommendation as they perfectly fit the cross-domain recommendation problem scenarios. With shared user identities among domains, it is easy to find corresponding users among different domains and form cross-domain recommendation tasks. In general, there are three representative datasets with multiple domains: \emph{Amazon}\cite{amazon}, \emph{Douban}\cite{douban} and \emph{Epinions}\cite{epinions}. 

\begin{itemize}
\item The \emph{Amazon} dataset is crawled from Amazon, the world’s largest e-commerce platform, which has the most extensive behavior data volume and good diversity. This dataset contains information of $24$ different domains (e.g., CDs and Vinyl, Electronics, Movies, and TV), depending on the type of items. For each interaction, a user gives an explicit rating score on a scale of $[1,5]$ to an item indicating the user's preference for this item. At the same time, this dataset also includes users' review texts on items, the interaction timestamps, and items' categories. 
\item The \emph{Douban} dataset is crawled from Douban, a renowned Chinese online social network, where users give ratings to three types of items: movie, music, and book. This dataset consists of three domains, that is, Douban-Movies, Douban-Music, and Douban-Books. Apart from ratings, this dataset also includes users' profiles involving gender, age, place of residence, tag, etc., and the attributes of items such as names and ownerships. In addition, social relationships among users are also available.
\item The \emph{Epinions} dataset is extracted from Epinions, a popular product review website, providing 587 distinct domains (e.g. Books, Videos and DVDs, Baby Care) and sub-domains. A user offers both an explicit rating score on a scale of $[1, 5]$ and reviews to an item in each interaction. Additionally, this dataset compromises user profiles including name, location, top rank, etc., and features of items like name, category, and description. More importantly, trust statements and "experts", i.e., category leads, top reviewers, or advisors,  are also available in this dataset. A user scores other users positive trust statement, i.e. $1$ when trusting their reviews, otherwise negative one, i.e. $0$ and "experts" offer more reliable trust statements. 
\end{itemize}

Typically, datasets with multiple domains can generate two different kinds of cross-domain recommendation tasks. In most cases, any two domains can be used to generate a cross-domain recommendation task (e.g., Amazon-Movies and TV $\leftrightarrow$ Amazon-CDs and Vinyl, Douban-Movies $\leftrightarrow$ Douban-Music) where users partially/fully overlap, and items no-overlap. Alternatively, some domains together with domains from other datasets can form another kind of cross-domain recommendation tasks (e.g., Amazon-Movie and TV $\leftrightarrow$ Douban-Movies, Amazon-Books $\leftrightarrow$ Douban-Books) where users no-overlap, and items partially/fully overlap.

\begin{table*}[t]
	\caption{A summary of datasets and corresponding papers}
	\label{Datasets summary}
	\centering
	\setlength{\tabcolsep}{1 mm}
	\begin{tabular}{c|c|c|l}
	\toprule
	\textbf{Dataset Type}&\textbf{Domain Type}& \textbf{Name of Dataset} & \textbf{Paper}\\
	\cline{1-4}
	\multirow{5}{*}{\shortstack{Datasets\\with\\Multiple\\Domain}}&\multirow{2}{*}{-} & \multirow{2}{*}{Amazon} & \cite{DBLP:journals/dss/ZhangWLLZ17,DBLP:conf/aistats/IwataK15,DBLP:conf/aaai/FuPWXL19,DBLP:conf/dasfaa/WangPWYFH18,DBLP:conf/cikm/KangHLY19,DBLP:conf/sigir/ZhaoLXDS20,DBLP:conf/recsys/CuiWZZ20,DBLP:conf/dasfaa/ZhaoHWH18,DBLP:conf/cikm/HuZY18,DBLP:journals/ijon/TanBQCC14,DBLP:conf/cikm/ZhaoLF19,DBLP:conf/ecir/LoniSLH14,DBLP:conf/cikm/WangZZWZH21,DBLP:conf/dasfaa/SongPWFHY17,DBLP:journals/corr/abs-2202-04893}\\
	& & & \cite{DBLP:journals/tkdd/MirbakhshL15,DBLP:journals/ijon/HuangZWHC19,DBLP:conf/www/HuCXCGZ13,DBLP:conf/dasfaa/LiPWXYH20,DBLP:conf/dasfaa/YanZZWLS20,hu2018mtnet,DBLP:conf/ijcai/ZhangLHMCLT20,DBLP:conf/ijcai/YuanYB19,DBLP:conf/wsdm/WangFGCH19,DBLP:conf/cikm/LiuLLP20,DBLP:conf/www/LiuZZLSX0X20,DBLP:journals/corr/abs-2202-04920,DBLP:conf/cikm/XuXCZ21,DBLP:journals/corr/abs-2104-08490,DBLP:conf/sigir/ZhuGZXXZL021,DBLP:conf/wsdm/ZhuTLZXZLH22}\\
	\cline{2-4}
	&\multirow{2}{*}{-} & \multirow{2}{*}{Douban} & \cite{DBLP:conf/cikm/ZhuC0LZ19,DBLP:conf/ijcai/ZhuWCLZ20,DBLP:conf/ijcai/XinLLHWG15,DBLP:conf/icdm/HeLZLNH18,DBLP:conf/aaai/ZhaoPXZLY13,DBLP:conf/aaai/ZhangJLD016,DBLP:conf/cikm/YangHQXW15,DBLP:conf/www/LianZXS17,DBLP:journals/tnn/ZhangLWZ19}\\
	& & & \cite{DBLP:conf/ecai/0001LWHWZ20,DBLP:conf/ijcai/MaZWLLM18,DBLP:conf/aaai/JiangCYXY16,DBLP:journals/corr/abs-2202-04920,DBLP:conf/sigir/ZhuGZXXZL021,DBLP:journals/corr/abs-2108-07976,DBLP:conf/wsdm/ZhaoYY22,DBLP:conf/ijcai/ManSJC17,DBLP:conf/ijcai/ZhuWCLOW18,DBLP:journals/corr/abs-2202-04893}\\
	 \cline{2-4}
	&- & Epinions & \cite{DBLP:conf/pkdd/RafailidisC16,DBLP:conf/cikm/MaYLK08a,DBLP:conf/wsdm/ShuWTWL18,DBLP:conf/ecai/0001LWHWZ20,DBLP:conf/cikm/RafailidisC17,DBLP:journals/tkdd/MirbakhshL15} \\
	 \cline{1-4}
	\multirow{15}{*}{\shortstack{Datasets\\with\\Single\\Domain}}&\multirow{4}{*}{Movie} & \multirow{2}{*}{MovieLens} &\cite{DBLP:journals/dss/ZhangWLLZ17,DBLP:conf/ijcai/LiYX09,DBLP:conf/pkdd/GaoLCLGG13,DBLP:conf/aistats/IwataK15,DBLP:conf/wsdm/HeZYY18,DBLP:conf/icml/LiYX09,DBLP:conf/um/ShiLH11,DBLP:conf/fuzzIEEE/HaoZL16,DBLP:conf/ijcai/ManSJC17,DBLP:conf/ijcai/ZhuWCLOW18,DBLP:conf/cikm/ZhuC0LZ19,DBLP:conf/ijcai/ZhuWCLZ20,DBLP:conf/dasfaa/ZhaoHWH18,DBLP:conf/icdm/HeLZLNH18,DBLP:journals/ijon/TanBQCC14,DBLP:conf/ecweb/EnrichBR13,DBLP:conf/recsys/Fernandez-TobiasC14,DBLP:conf/www/GaoCFZ00J19}\\
	&& & \cite{DBLP:conf/icml/LiuHW15,DBLP:journals/tnn/ZhangLWZ19,DBLP:journals/ijon/HuangZWHC19,DBLP:conf/icdm/FangGLLL15,DBLP:conf/kdd/AgarwalCL11,DBLP:journals/tcyb/LiZLZ15,DBLP:conf/aaai/PanXLY10,DBLP:conf/aaai/ZhangJLD016,DBLP:conf/cidm/KumarKHCA14,DBLP:conf/aaai/PanXY12,DBLP:conf/aaai/RenGLG15,DBLP:conf/www/LianZXS17,DBLP:conf/kdd/ChenHL13,DBLP:conf/aaai/PereraZ20,DBLP:journals/corr/abs-2108-07976}\\
	\cline{3-4}
	& & Netflix & \cite{DBLP:journals/dss/ZhangWLLZ17,DBLP:conf/aistats/IwataK15,DBLP:conf/cikm/MorenoSRS12,DBLP:conf/ijcai/ManSJC17,DBLP:conf/ijcai/ZhuWCLOW18,DBLP:conf/aaai/ZhaoPXZLY13,DBLP:conf/aaai/ZhangJLD016,DBLP:conf/ijcai/PanLXY11,DBLP:conf/ijcai/LiZLZXW11,DBLP:conf/www/GaoCFZ00J19,DBLP:conf/aaai/PanXY12,DBLP:journals/tnn/ZhangLWZ19,DBLP:journals/tcyb/LiZLZ15,DBLP:conf/kdd/SinghG08,DBLP:conf/sdm/LuPXYZZ13,DBLP:conf/aaai/PanXLY10}\\
	 \cline{3-4}
	& & EachMovie & \cite{DBLP:conf/ijcai/LiYX09,DBLP:conf/pkdd/GaoLCLGG13,DBLP:conf/aistats/IwataK15,DBLP:conf/wsdm/HeZYY18,DBLP:conf/icml/LiYX09,DBLP:conf/aaai/ZhangJLD016,DBLP:conf/aaai/RenGLG15,DBLP:conf/um/BerkovskyKR07}\\
	 \cline{2-4}
	&\multirow{2}{*}{Music} & Yahoo-Music &\cite{DBLP:journals/dss/ZhangWLLZ17,DBLP:conf/kdd/LiL14} \\
	\cline{3-4}
	& & Last.FM& \cite{DBLP:conf/ijcnn/ZhangH0019} \\
	
	\cline{2-4}
	& \multirow{2}{*}{Book} & Book-Crossing & \cite{DBLP:conf/ijcai/LiYX09,DBLP:conf/pkdd/GaoLCLGG13,DBLP:conf/wsdm/HeZYY18,DBLP:conf/icml/LiYX09,DBLP:conf/cidm/KumarKHCA14,DBLP:conf/aaai/RenGLG15} \\
	\cline{3-4}
	&& Library Thing & \cite{DBLP:journals/dss/ZhangWLLZ17,DBLP:conf/um/ShiLH11,DBLP:conf/fuzzIEEE/HaoZL16,DBLP:conf/recsys/Fernandez-TobiasC14,DBLP:conf/kdd/ChenHL13} \\
	
	\cline{2-4}
	& \multirow{4}{*}{\shortstack{Social \\ Interaction}} & Twitter &\cite{DBLP:conf/sigir/Wang0NC17,DBLP:conf/aaai/PereraZ20,DBLP:conf/mm/PereraZ17,DBLP:conf/mir/YanSX15}\\
	\cline{3-4}
	&& Youtube & \cite{DBLP:conf/ijcai/PereraZ18,DBLP:conf/aaai/PereraZ20,DBLP:conf/mm/PereraZ17,DBLP:conf/mir/YanSX15} \\
	\cline{3-4}
	&& Weibo & \cite{DBLP:conf/cikm/YangHQXW15,DBLP:conf/ijcai/MaZWLLM18,DBLP:conf/aaai/JiangCYXY16} \\
	\cline{3-4}
	&& DBLP & \cite{DBLP:conf/icml/LiuHW15} \\
	\cline{2-4}
	&\multirow{3}{*}{\shortstack{Point\\of\\Interest}} & Yelp &\cite{DBLP:conf/recsys/SahebiB15,DBLP:conf/ecir/ManotumruksaRMO19,DBLP:conf/sigir/KrishnanDBYS20} \\
	\cline{3-4}
	& & Brightkite &  \cite{DBLP:conf/ecir/ManotumruksaRMO19}\\
	\cline{3-4}
	& & Foursquare & \cite{DBLP:conf/ecir/ManotumruksaRMO19} \\
	
	 \cline{2-4}
	 &Others & Cheetah Mobile & \cite{DBLP:conf/cikm/HuZY18,DBLP:conf/dasfaa/YanZZWLS20,hu2018mtnet} \\
	\bottomrule
	\end{tabular}
\end{table*}

\subsection{Datasets with Single Domain}
\subsubsection{Movie}
Movie datasets are favored by a majority of works because they have innate shared movies in different categories which can be treated as different domains to form cross-domain recommendation tasks. Also, as many movies in these datasets are in common, a recommendation scenario based on overlapping items can generate.There are three widely used movie datasets: \emph{MovieLens}~\cite{movielens}, \emph{Netflix}~\cite{netflix} and \emph{EachMovie}~\cite{eachmovie}.

\begin{itemize}
\item The \emph{MovieLens} dataset is crawled from MovieLens, a well-received movie recommendation website, which contains a group of datasets leveled by a rating scale. Among them three stable benchmark datasets, i.e., MovieLens-100K, MovieLens-1M and MovieLens-20M, are most popular. The differences between them are that they have a distinct volume of interactions. Each interaction in this dataset includes ratings on a scale of $[0.5, 5]$ with half-rating increments, attributes of movies, and tags user-generated. Movies are separated into 18 distinct genres including Action, Adventure, Horror, and so on.

\item The \emph{Netflix} dataset, offered by Netflix for a competition, consists of about $100,000,000$ ratings for $17,770$ movies given by $480,189$ users. Each interaction consists of ratings on a scale of $[1, 5]$, the interaction time, and attributes of movies.

\item The \emph{EachMovie} dataset is extracted from EachMovie recommendation service compromising $2,811,983$ ratings from $1$ to $6$ entered by $72,916$ users for $1628$ different movies. 
\end{itemize}

\subsubsection{Book.}
The cross-domain book recommendation is another prevalent task. Book-Crossing~\cite{bookcrossing} and LibraryThing~\cite{librarything} are two majorly adopted datasets for the task.

\begin{itemize}

\item The \emph{Book-Crossing} dataset is crawled from the Book-Crossing community contains $278,858$ users with $1,149,780$ ratings about $271,379$ books. Interactions in this dataset compromise explicit ratings on a scale of $[1, 10]$ to books. Demographic information about users including location and age are provided while content-based information of each book, i.e., tile, author, year of publication, and publisher, are given.

\item The \emph{Library Thing} dataset is collected from Library Thing, an online book review website, which publishes information about books and enables users to create their virtual libraries and tag books. It consists of $2,056,487$ tuples with $7,279$ users, $37,232$ books, and $10,559$ tags. In each dataset, a user gives ratings ranged between $1$ and $5$, reviews, tags to a book. In addition, friend relations among users are given which are similar to the trust relations of social networks in Epinions.

\end{itemize}

Although user identities are not identical across different datasets, each book in these datasets is labeled by a unique ISBN ID. It is possible to match the same book in different domains to build cross-domain recommendation scenarios where users do not overlap and items are fully or partially overlapped.

\subsubsection{Music.}

Music datasets record user-music interactions and Yahoo! Music~\cite{yahoo} and Last.FM~\cite{lastfm} are two representative music datasets.

\begin{itemize}
\item The \emph{Yahoo! Music} dataset, offered by Yahoo Research, consists of a wealth of information about music. Users give explicit ratings to entities of four different types, i.e., tracks, albums, artists, and genres, constituting four types of interactions. In addition to ratings, each interaction also contains attributes of entities.
\item The \emph {Last.FM} is released by HetRec in $2011$. Different from previously introduced datasets, this dataset contains the number of times a song has been listened to by a user instead of explicit ratings as well as tags generated by the user.
\end{itemize}

\subsubsection{Social Interaction.}
In most instances, social interaction datasets are used as auxiliary information to assist cross-domain recommendation. Twitter~\cite{twitter}, Youtube~\cite{youtube}, Weibo~\cite{weibo} and DBLP~\cite{dblp} are four typically used datasets.

\begin{itemize}
\item The \emph{Twitter} dataset is crawled from one of the most widespread social networks, Twitter. Each interaction includes tweet content, the interaction time, and tags users generated.
\item The \emph{Youtube} dataset, given by Youtube, consists of user-video interactions provided with video ID, video title, label, class, the interaction time, and detailed description.
\item The \emph{Weibo} dataset is crawled from the largest Chinese Twitter Weibo. Each tweet contains tweet content, user identification, interaction time, comments, and tags. Each user owns a profile of gender, name, followers, and followees, which is possible to form a social relation network.
\item The \emph{DBLP} dataset is offered by a famous citation network DBLP. Each interaction compromises paper, abstract, authors, year, venue, title, and citations. Moreover, citations of papers are used to construct a social relation network.
\end{itemize}

\subsubsection{Point-of-Interest.}

Point-of-interest datasets record the behaviors of users at the Point-of-Interest (POI) which are widely used for cross-domain venue recommendation. The most popular POI datasets include \emph{Yelp}~\cite{yelp} and two check-in datasets, \emph{Brightkite}~\cite{brightkite} and \emph{Foursquare}~\cite{foursquare}.

\begin{itemize}
\item The \emph{Yelp} dataset is offered by the largest review site in the U.S., i.e., Yelp, for a challenge. This dataset consists of ratings on a scale of $[1,5]$, reviews, and tips of each interaction as well as social relation network.
\item The \emph{Brightkite} dataset is extracted from a location-based social networking service provider Brightkite, where users shared their locations by checking-in. Interactions with check-in time and location as well as friendship network are available in this dataset. 
\item The \emph{Foursquare} dataset is collected from Foursquare, a location data platform for understanding how people move through the real world. The dataset consists of check-in data for different cities accompanied by timestamps, GPS coordinates, and semantic meaning (represented by fine-grained venue categories).
\end{itemize}

\subsubsection{Others.}

\begin{itemize}
    \item The \emph{Cheetah Mobile} dataset is a mobile dataset collected from Cheetah. It consists of two domains, that is, app installation and news browsing, which makes it naturally fit for cross-domain recommendation scenarios where users overlap but items do not. Each interaction in the news browsing domain contains contents of news, the interaction time, and user profiles, while each interaction of app installation includes app IDs and some metadata about both users and apps.
\end{itemize}

\section{Future Directions and Challenges}
\label{challenge_future_direction}

While existing works have established a solid foundation for cross-domain recommendation research, there are still some further opportunities. In this section, we discuss some future research directions and the main challenges for them.
 
\subsection{Exploring Unstudied Recommendation Scenarios}
As we have introduced in Section~\ref{recommendation_scenario}, based on our proposed taxonomy, there exist $9$ different recommendation scenarios for cross-domain recommendation. However, three of them (i.e., user partial overlap $\&$ item partial overlap, user partial overlap $\&$ item full overlap, and user full overlap $\&$ item partial overlap) have not been studied so far. These three recommendation scenarios all assume that both the users and the items of two domains overlap with each other.
With the increase in the number of service platforms, more and more users begin to interact on multiple platforms with similar functions, which makes the generation of such datasets possible. For example, the same users watch videos on Tencent and iQIYI, which makes user sets of these two platforms overlap. At the same time, there are the same videos (i.e., the items) provided by these two platforms. Therefore, a dataset supporting these kinds of studies can be formed. The same phenomenon also exists on e-commerce platforms like Taobao and Amazon. Therefore, future researches on cross-domain recommendation can be conducted under these unexplored recommendation scenarios.
The main challenge in performing cross-domain recommendation in these scenarios lies in designing customized knowledge transfer mechanisms so that cross-domain knowledge can be transferred through both overlapping users and overlapping items. Existing knowledge transfer mechanisms are designed either only for overlapping users or only for overlapping items, but not for both. Because users and items have different characteristics and contain interrelated information, it is difficult to obtain satisfactory performance by directly applying existing knowledge transfer mechanisms.

\subsection{Adopting the Latest Advances in Deep Learning}

The cross-domain recommendation problem can be regarded as a sub-problem of the recommendation system, which is closely related to the traditional recommendation problem. Therefore, the methods proposed in traditional recommendation problems, especially the ways of extracting knowledge in each domain, can be directly referred to and applied to cross-domain recommendation problems. Moreover, technological developments in other related fields can also be introduced. For example, technology in the field of natural language processing can be used to process user comments and item contents that act as auxiliary information, and technology in the field of computer vision can be used to process visual information of items~\cite{DBLP:conf/www/LiuZZLSX0X20}. 
When adopting the latest advances in deep learning, e.g., meta-learning, for cross-domain recommendation, the main challenge lies in leveraging the advantages of these new advances to meet the unique characteristics of cross-domain recommendation.

In terms of specific methods, Hu et al.~\cite{hu2018mtnet} utilized a \textit{memory network} to attentively extract useful information to be transferred from text content. 
Hao et al.~\cite{DBLP:conf/kdd/HaoLXGTZL21} proposed an Adversarial Feature Translation (AFT) framework. The generator aims to generate item candidates in all domains that users may click and the discriminator aims to distinguish user's real clicked items from the fake clicked items provided by the generator.
Li et al.~\cite{DBLP:conf/wsdm/LiZZYCSKN22} also proposed an \textit{adversarial learning} framework in which the generator produces users' domain-independent embeddings and the discriminator is trained to make user embeddings in the two domains become statistically indistinguishable. 
Zhu et al.~\cite{DBLP:conf/wsdm/ZhuTLZXZLH22} proposed to incorporate the strengths of \emph{meta-learning} into cross-domain recommendation and learned a personalized bridge mapping function for each user in the EMCDR-based framework~\cite{DBLP:conf/ijcai/ManSJC17}. 
However, with the rapid development of deep learning and neural networks, there are still many new technologies and structures that can be utilized to improve the performance of cross-domain recommendation.

\subsection{Exploring Robustness of Recommendation}

Robustness is critical to the generalization of models. If the robustness of a model is poor, a little disturbance may greatly damage the prediction accuracy. 
There are two main challenges in achieving robustness of models in cross-domain recommendation. First, it is a great challenge to transfer useful knowledge between domains while avoiding negative transfer. The transferred knowledge may contain noise and hence reduce the robustness of the models. Second, cross-domain recommendation models tend to be more complex because they consider not only extracting useful knowledge within each domain but also transferring such knowledge across domains. Complex models are more susceptible to noise interference and have lower robustness.
There have been several studies on the robustness of cross-domain recommendation models. Yan et al.~\cite{DBLP:conf/dasfaa/YanZZWLS20} proposed a new Adversarial Cross-Domain Network (ACDN) which is based on the similar framework of CoNet. When learning parameters, ACDN adds intentional perturbations on the embedding representations to generate adversarial examples and help to learn robust parameters. Zhang et al.~\cite{DBLP:journals/dss/ZhangWLLZ17,DBLP:journals/tnn/ZhangLWZ19} aimed at the "negative transfer" problem and applied domain adaptation functions to ensure the consistency of the transferred knowledge, which can enhance the robustness of the models. 
However, these are only preliminary explorations, and there is still a large research space on the robustness of cross-domain recommendation models.

\subsection{Scalability of Deep Cross-domain Recommendation Models}

The scalability of recommendation systems refers to whether the proposed models can perform recommendations effectively and timely when facing large-scale datasets. Model scalability is attracting more and more attention which is because, with the rapid development of informatization and digitalization, the amount of information grows exponentially. For recommendation systems, more and more users register and use a service or system, a large number of new products or items are produced and released every day, and tens of billions of user logs are generated every moment. All these components lead to the generation of a large volume of huge datasets, which presents a great challenge to the scalability of the proposed recommendation models when applied to real-world applications. 
For cross-domain recommendation, there are two main challenges in designing scalable models of cross-domain recommendation. On the one hand, compared with single-domain recommendation models, cross-domain recommendation models generally have more complex structures as they need to transfer knowledge across domains. On the other hand, cross-domain recommendation integrates information from multiple domains. This indicates that the size of datasets becomes much larger than that of datasets in single-domain recommendation. Therefore, cross-domain recommendation typically requires longer training time.

\subsection{Explainability of Cross-Domain Recommendation}

In recent years, the explainability of recommendations has become a hot research topic. Studies on the explainability of recommendations generally fall into two categories. The first category of studies focuses on the explainability of recommendation results, that is, explaining why items are recommended to a specific user. Metadata such as user profiles, textual contents, and reviews are widely utilized. Some approaches directly highlight the most important words or phrases in the related reviews to explain the recommendations. Other approaches rely on specially designed language models to generate textual explanations. Knowledge graphs are also broadly exploited as they can establish multi-hop relations between users and items. They can directly give explanations, for example, a recommended movie is starred by the same actor from another movie the user has watched. The second category of studies focuses on the explainability of models themselves, that is, how the models generate the recommended items. Due to the capability to pinpoint and amplify salient features that greatly affect the recommendation, attention mechanisms are regarded as a reasonable and reliable way to explain the decision-making procedure in many approaches~\cite{DBLP:conf/ijcai/YingZZLXXX018}. By visualizing the weights of the attention mechanisms, it can explicitly indicate which part of the information is more important or which step of inputs is more strongly associated with the outputs. Moreover, some dimensionality reduction methods, such as T-Distributed Stochastic Neighbor Embedding (T-SNE), map the learned higher-dimensional latent representations to lower-dimensional space (i.e., 2D or 3D spaces), are also used to interpret the intermediate results of models.

In the context of cross-domain recommendation, however, there still exist two main challenges in studying the explainability of models. First, from the perspective of explainability of recommendation results, due to the differences between domains, the metadata, especially the textual contents and reviews, may use different languages and have different styles. This may impair the readability of the generated explanations as the generated explanations will include both within-domain information and cross-domain information. Second, from the perspective of explainability of models themselves, as cross-domain recommendation models need to perform knowledge transfer between domains, many additional factors, such as within-domain knowledge, cross-domain knowledge, domain-specific representations, and domain-invariant representations, should be considered when improving the model explainability.
We believe if the abovementioned challenges could be addressed, cross-domain models with explainability would be beneficial to improve the transparency, persuasiveness, and trustworthiness of cross-domain recommendation.

\section{Conclusion}
\label{conclusion}
In this paper, we provide a comprehensive and systematic investigation on cross-domain recommendation, which is a powerful tool to solve the data sparsity and cold-start problems in traditional recommender systems. We first proposed a two-level taxonomy of cross-domain recommendation scenarios and recommendation tasks for organizing and clustering existing works. 
Under each research scenario, we systematically sort out and summarize existing research works in terms of methods being used.
Moreover, a detailed introduction about frequently used datasets including datasets with multiple domains and datasets with a single domain is provided.
Finally, we discuss some promising potential research directions for further research on cross-domain recommendation.
We hope that this survey can provide both newcomers and experts of cross-domain recommendation with a comprehensive understanding of the problem definition of this field, clarify existing works clearly, and shed some light on future studies.

\section*{acknowledgments}
This research is supported in part by the 2030 National Key AI Program of China 2018AAA0100503, National Science Foundation of China (No. 62072304, No. 61772341, No. 61832013, No. 62172277), Shanghai Municipal Science and Technology Commission (No. 19510760500, No. 21511104700, No. 19511120300), the Oceanic Interdisciplinary Program of Shanghai Jiao Tong University (No. SL2020MS032), Scientific Research Fund of Second Institute of Oceanography, the open fund of State Key Laboratory of Satellite Ocean Environment Dynamics, Second Institute of Oceanography, MNR, GE China, and Zhejiang Aoxin Co. Ltd.

\newpage


\bibliographystyle{ACM-Reference-Format}

\end{document}